\documentclass[aps,prl,reprint,groupedaddress]{revtex4-2}
\usepackage{graphicx}

\begin{document}


\title{Charge-Order pattern in $\kappa$-BEDT-TTF salts:\\
A model independent, selection rule approach}


\author{Alberto Girlando}
\affiliation{Molecular Materials Group (MoMaG), Parma, Italy}
\email{alberto.girlando@momag.it}
\homepage{\\albertogirlando@wordpress.com}

\date{\today}

\begin{abstract}
A well known vibrational selection rule is invoked to determine the
charge order (CO) pattern displayed by the most interesting members of the widely
studied class of $\kappa$-BEDT-TTF salts and their molecular analogues.
The selection rule indicates that an easily identifiable Raman band should be present if intra-dimer charge disproportionation 
occurs, constituting a sort of model independent ``litmus test''.
The absence of such a band in the available Raman spectra of
$\kappa$-BEDT-TTF salts exhibiting static of fluctuating CO, strongly
suggests that the CO pattern involves inter-dimer CO, in contrast with the universally accepted intra-dimer disprortionation.
This paper therefore poses a challenge for the theoretical modeling of the concerned salts: either the lack
of the Raman band sensitive to intra-dimer CO pattern is explained,
or the modeling of the multivariate, unique physical properties
 of these $\kappa$-phase salts (spin frustration, electronic ferroelectricity etc.) has to start from the perspective of a different ground state.
\end{abstract}


\maketitle

\textit{Introduction}---Symmetry breaking is a cornerstone of solid state and materials physics, providing a unified framework for the understandings
of complex quantum processes. This is well illustrated
in the field low-dimensional organic salts, where the
interplay of charge, spin and phonon degrees of freedom yield
many broken-symmetry electronic ground states and related quantum phenomena, such as superconductivity, electronic ferroelectricity, spin liquid states and so on \cite{dressel20}. In particular, unconventional superconductivity and electronic ferroelectricity in 
bis(ethylenedithio)tetrathiafulvalene (BEDT-TTF, or ET for short) salts appears to be strongly interconnected with charge-order (CO) symmetry breaking \cite{clay19,lang25}    

Precise knowledge of the actual ground state and CO pattern
is of course a necessary prerequisite for appropriate modeling
of the involved phenomena. Standard x-ray analysis is important,
but under certain circumstances may not be enough. For instance, 
it probes long distance ordering, underestimating disorder
and fluctuations \cite{canossa21,salzillo21}. Moreover, the refinement
process requires the initial assumption of a structure, and
assessing the presence/absence of the inversion center symmetry may be challenging. 

It is well known that vibrational spectroscopy, notably
Raman spectroscopy in the lattice phonons region, can be used as a structural tool to complement the x-ray analysis \cite{brillante08,girlando16} Here, however,
I want to stress the use of an elementary, but very stringent
symmetry-based selection rule - the so-called ``mutual exclusion''
rule - to discriminate, independently from any predetermined
assumption, the static or fluctuating CO pattern exhibited by
some $\kappa$-phase ET salts. 
  
\textit{Four basic structural patterns}---In dealing with molecular Charge-Transfer (CT) crystals it is useful to consider, 
besides the space group symmetry, also
the molecular symmetry and the site symmetry \cite{turrell72}. Due to the CT interaction, in these crystals the generally planar
molecules arrange in quasi one-dimensional (1D) or two-dimensional (2D) packing, and the molecular totally symmetric
$a_g$ vibrations may be coupled to the
CT transition (e-mv coupling), giving rise to peculiar spectroscopic effects \cite{painelli86}. The site symmetry
is instead the symmetry at a given site of the crystal,
where a molecule can reside or not. In the former case,
it is a subgroup of the molecular symmetry.  

Molecular crystals generally belong to the monoclinic, low symmetry groups, and the most relevant site symmetry is the
inversion center. Thus the rule of mutual exclusion applies:
Raman active modes are cannot be infrared (IR) active and vice-versa. This selection rule is very stringent, and if
the packing deviates from the inversion center symmetry,
also the converse is true: all the modes become both Raman
and IR active.
\begin{figure} [hb]
	\centering
	\includegraphics[width=0.67\linewidth]{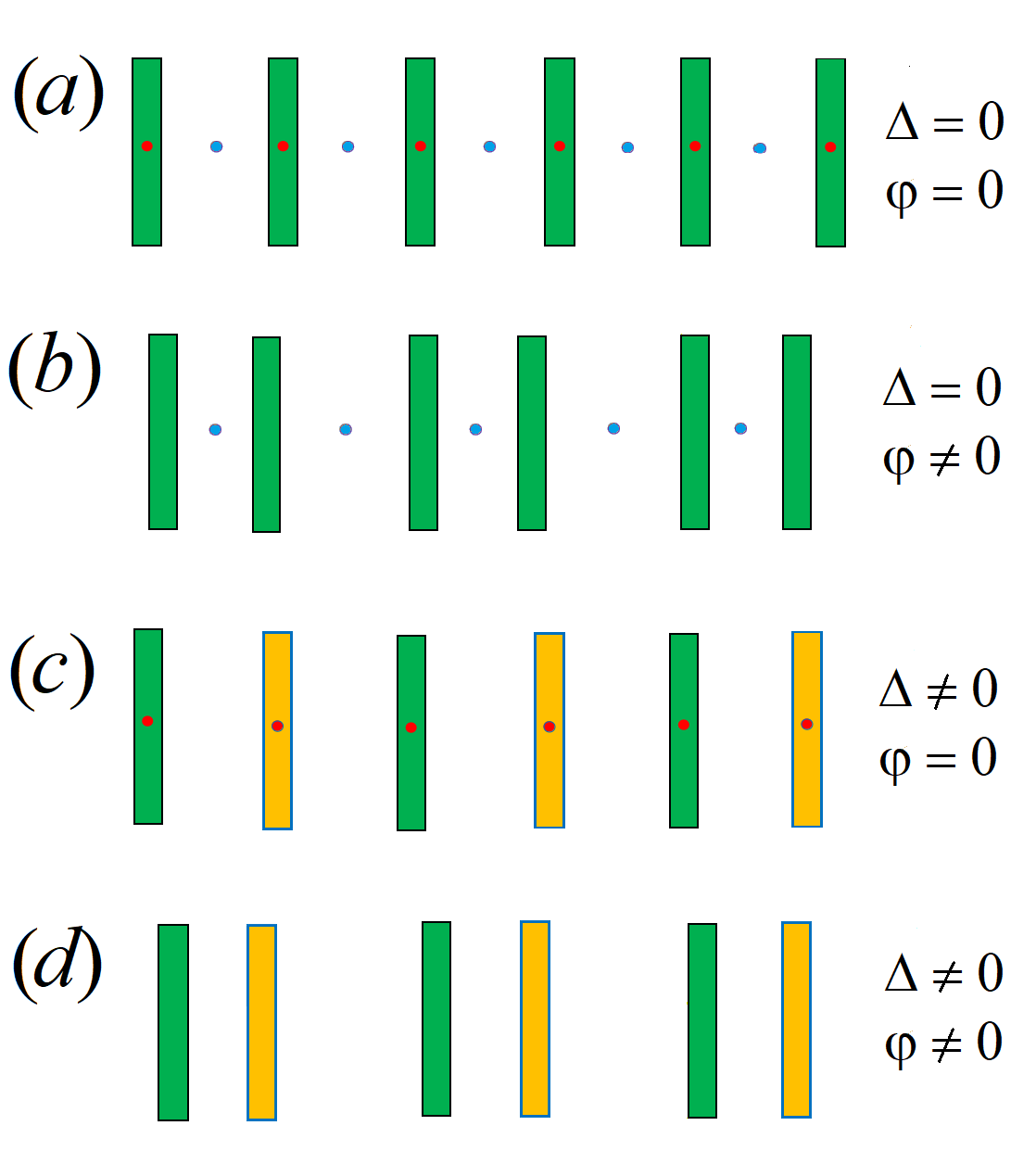}
	\caption{Four basic 1D CT stack patterns: a) Equally spaced 
		identical molecules (represented by rectangles); b) Alternating different nearest-neighbor distances between
		equal molecules; c) Equally spaced electron donor (yellow rectangle) and acceptor molecules (green rectangle); d)
		Alternating distances between different molecules.}
	\label{fig:4stacks}
\end{figure}

 In order to illustrate with practical examples the consequences of the mutual exclusion selection rule on the vibrational spectra, and how it is sensitive to the crystal packing, I will consider four types of 1D stacks (Figure \ref{fig:4stacks}) \cite{pecile89,bozio91,tomic15}. What follows can easily extended
 to other crystal packings, if needed. In the framework of the Hubbard model, the stacks are distinguished by the difference
 between the nearest-neighbor CT integrals ($\varphi = t - t'$)
 and on-site energy ($\Delta$). In Figure \ref{fig:4stacks}(a)
 the stack is made up by equally spaced identical
 molecules, $\Delta=0$ and $\varphi = 0$: there is an inversion center on the molecular sites (red dots) and between the
 molecular sites (cyan dots). The mutual exclusion rule holds in full: there is no band coincidence between Raman and IR,
 and no effects due to e-mv coupling since $a_g$ molecular
 modes are decoupled from the CT transition
 \cite{painelli86}.
 
 One-dimensional systems are subject to the Peierls instability
 yielding a distortion of the stack, represented 
 in Figure \ref{fig:4stacks}(b) for the case of stack dimerization,
 as it occurs in the half-filled case. The inversion center
 on the molecular sites is lost, and the stack can
 be considered as made up of weakly interacting dimers, $\Delta=0$ and $\varphi \neq 0$. Each dimer,
taken as a super-molecule, resides on inversion center, so the
mutual exclusion selection rule holds, and there is no
Raman-IR frequency coincidence. However, the vibrations of
each molecule couple in-phase (i-ph) and out-of-phase (o-ph). The i-ph $a_g$ phonons are Raman active, and again decoupled from
the CT transition, but the o-ph ones cause a charge oscillation back and forth within the dimer. Their consequent strong interaction with the CT electron causes a frequency lowering and intensity borrowing from the nearby CT transition, so the e-mv coupled $a_g$ phonons
appear with strong intensity in the IR spectra polarized perpendicular to the stack \cite{painelli86}.
\begin{figure}
	\centering
	\includegraphics[width=0.8\linewidth]{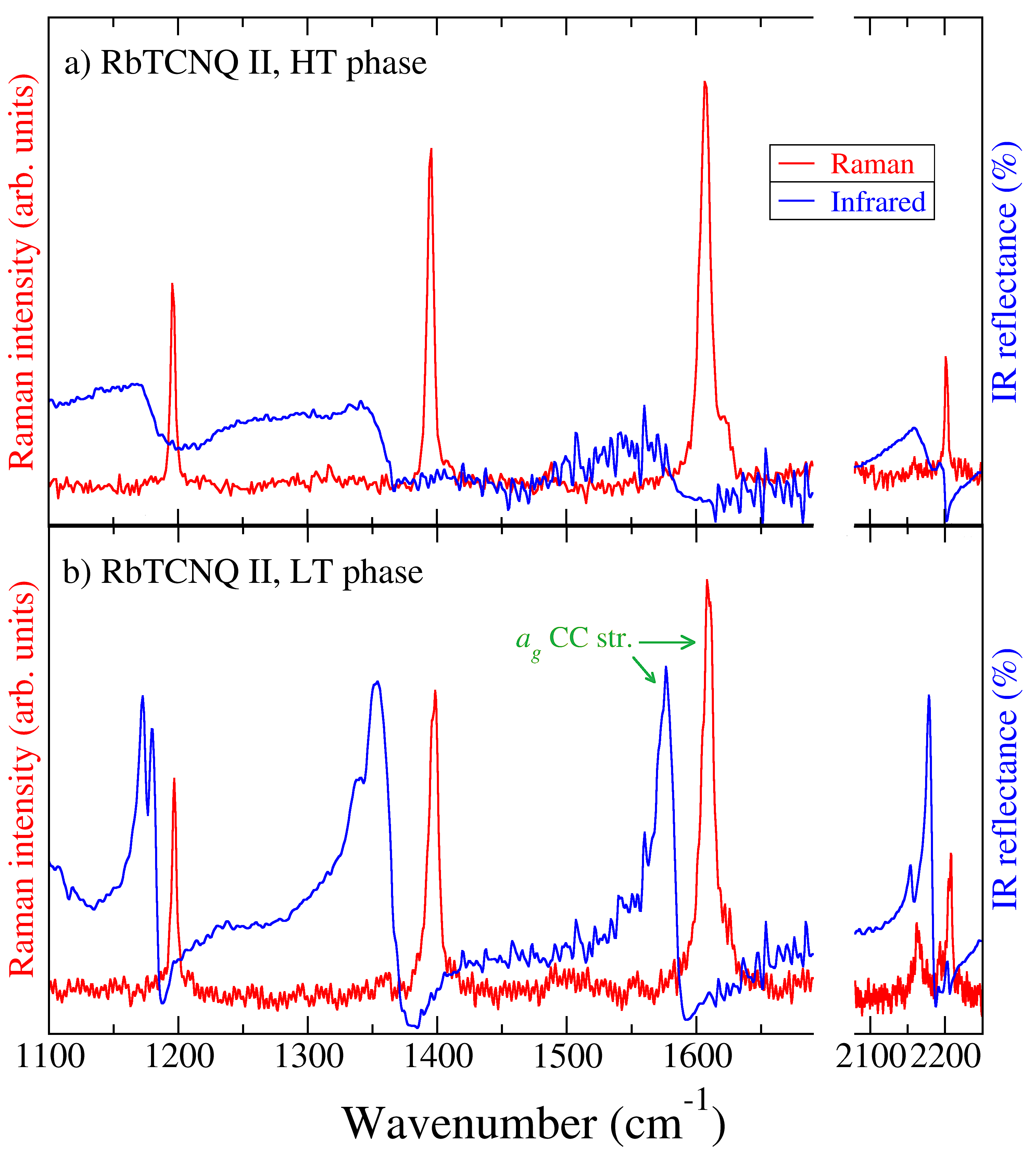}
	\caption{Top panel, a-type stack: RbTCNQ(II) at 360 K. Bottom panel, b-type stack, RbTCNQ(II) at 200 K. Raman (red line) and IR reflectance (blue line) spectra are overlapped in the same panel and scaled to display comparable intensities.
	IR reflectance spectra polarized perpendicular to the stack.}
	\label{fig:RbTCNQrir}
\end{figure}

Alkaline TCNQ salts provide a good example of the use of vibrational spectroscopy to characterize the Peierls
transition in half-filled systems \cite{bozio97}.
Figure \ref{fig:RbTCNQrir} reports the polarized Raman and IR
reflectance spectra of RbTCNQ(II) in the high temperature
(HT) and low temperature (LT) phases. The HT phase crystallizes
in the $P\overline{1}$  system, one molecule per unit cell,
so is representative of class (a) stack (Figure \ref{fig:4stacks}). The mutual exclusion rule holds, and the modes are not coupled to the CT electron: There is no strong IR band (blue line) corresponding to the $a_g$ Raman active bands (red line). By lowering the temperature below 250 K the stack
appears dimerized, as signaled by the insurgence of strong
e-mv induced IR bands associated with the o-ph coupling of
the $a_g$ phonons. Therefore, as shown in the bottom panel
of Figure \ref{fig:RbTCNQrir}, each $a_g$ Raman band has
a IR counterpart at lower frequency - in the Figure the
i-ph and o-ph coupled C=C stretching is signaled as one example.
  
It is worth mentioning that the presence of a real structural
dimerization transition has been questioned \cite{mcqueen09},
as x-ray analysis confirm the $P\overline{1}$ space group at 100 K, so the LT phase should be the realization of a Bond Order Wave (BOW) state. The present paper cannot discuss the merits of this
interpretation with respect to the alternative scheme of
a disorder-to-order phase transition. What I want to stress here is that the transition is very broad \cite{bozio97},
and that remnant of the IR e-mv induced bands are present
in the HT phase (top panel of Figure \ref{fig:RbTCNQrir}).
Vibrational spectroscopy probes short range order, as opposed
to x-ray, and is therefore sensitive to static or dynamic disorder, particularly if this involves the local loss of
the inversion center \cite{canossa21,salzillo21}.
\begin{figure}
	\centering
	\includegraphics[width=0.8\linewidth]{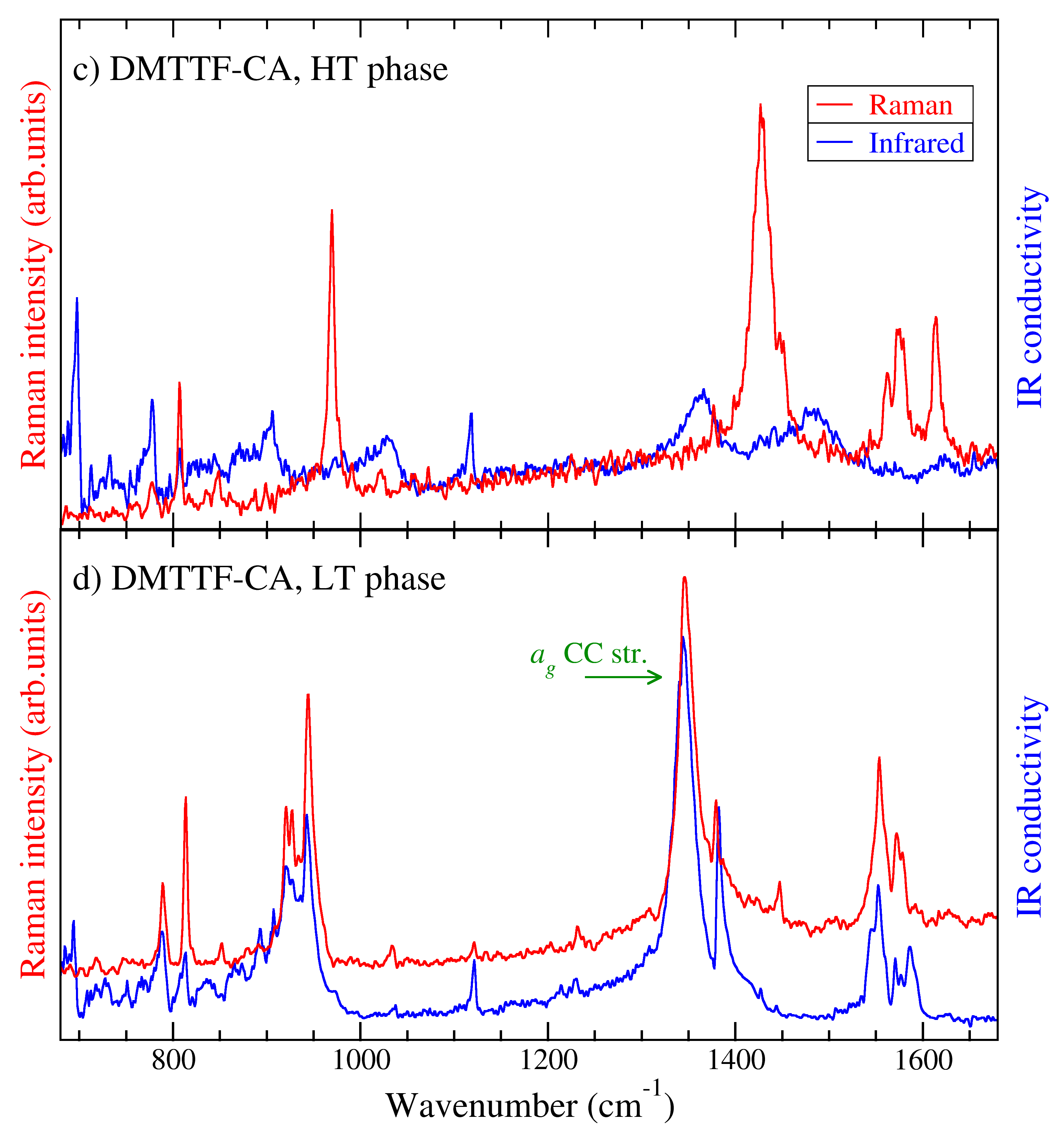}
	\caption{Top panel, c-type stack: DMTTF-CA at 300 K. Bottom panel, d-type stack at 40 K. Raman: red line; IR conductivity, polarized parallel to the stack: blue line. From Ref. \cite{ranzieri07a}.}
	\label{fig:DMTTF-CA}
\end{figure}

Donor-Acceptor CT crystals are examples of c-type stack,
as electron Donor and Acceptor molecules alternate with
equal distance along the stack. The Raman and IR spectra
of DMTTF-CA at room temperature are reported as an example
in the top panel of Figure \ref{fig:DMTTF-CA} \cite{ranzieri07a}. In this case the inversion center is on the molecular sites, and the mutual exclusion selection rule 
still holds. However, at variance with the HT phase of 
RbTCNQ(II) (Figure \ref{fig:RbTCNQrir}a) the molecular $a_g$
vibrations are coupled to the CT electron, and the Raman bands
are broadened and lowered in frequency with respect to
an hypothetic molecule with the same charge \cite{painelli86}.  

Also Donor-Acceptor CT crystals are subject to the Peierls
instability, yielding to dimerization of the stack, cf. Figure \ref{fig:4stacks}(d). In this case all the inversion centers
are lost, and all the vibrations become Raman and IR active
at the same frequency. The Raman, and the IR spectra polarized
parallel to the stack are dominated by the $a_g$ modes,
so the relative intensities are comparable,
as nicely shown in the bottom panel of Figure \ref{fig:DMTTF-CA}.
Here, the Raman and IR spectra of the low-temperature
dimerized phase of DMTTF-CA are shown, and the Raman (red) and
IR conductivity (blue) spectra are almost indistinguishable.

\textit{CO patterns in ET $\kappa$-phase salts}---
The concepts summarized in the previous Section have long been
known, perhaps they receded in the background with the advancement of
the field. In the following they will be applied to
determine the charge order (CO) pattern in the ET $\kappa$-phase
salts, the most intensively studied class of ET salts,
since it displays many intriguing phenomena. Among these,
unconventional  superconductivity, frustrated
quantum magnetism and ferroelectricity, all somehow related
to the CO, the uneven distribution of the charge among 
the ET units \cite{lang25}. Actually, the CO phase transition
occurs rarely in $\kappa$-phase salts, and a lot of attention
has been lately devoted to $\kappa$-(ET)$_2$Hg(SCN)$_2$Cl ($\kappa$-HgCl for short) CO transition,
where strong evidence was provided for electronic ferroelectricity \cite{drichko14,gati18,thomas19,hassan20,liebman26,urai26}.

\begin{figure}
	\centering
	\includegraphics[width=1.0\linewidth]{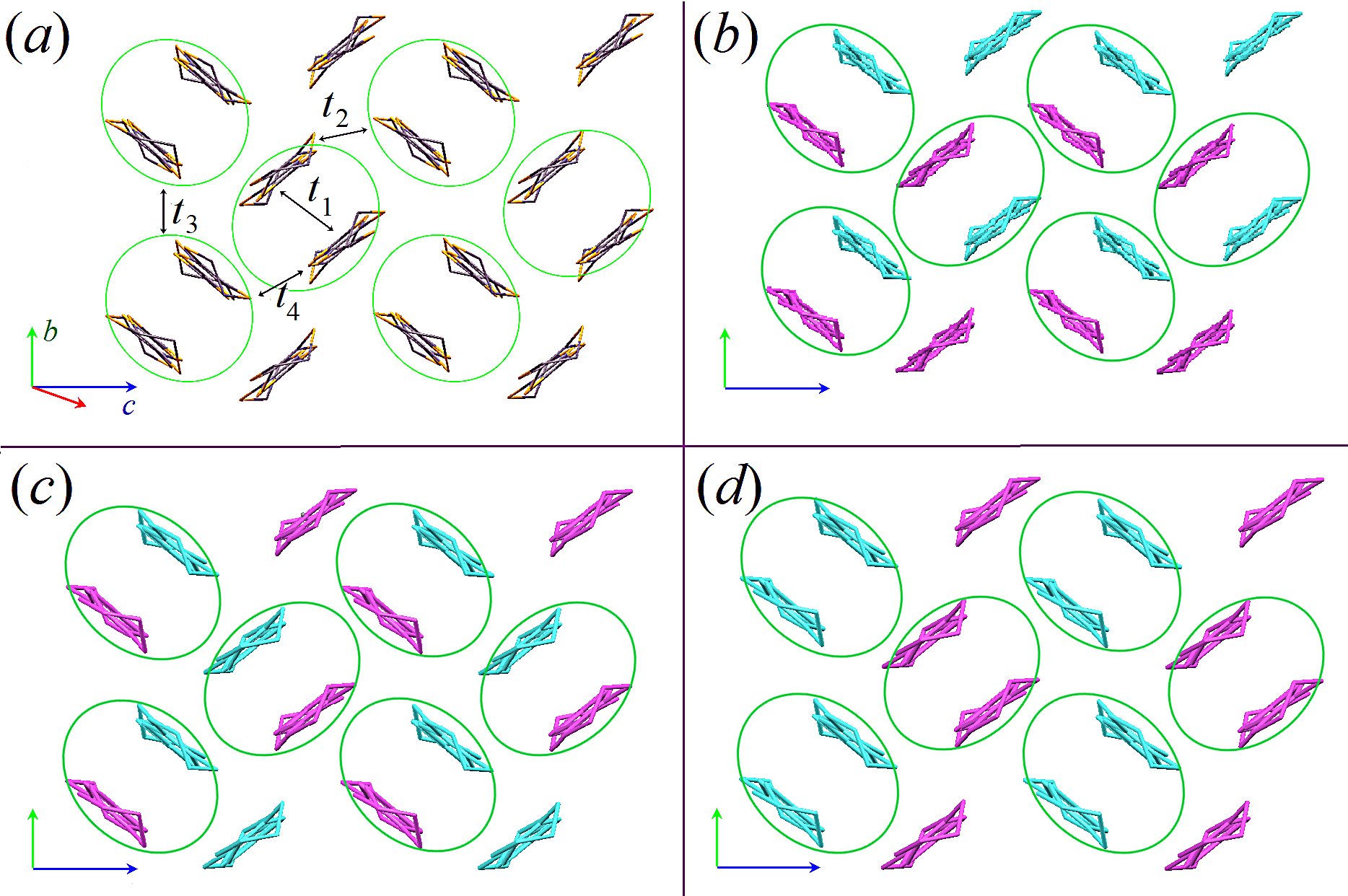}
	\caption{(a) ET layer in the metallic phase of $\kappa$-HgCl, viewed along the molecules’ long axes, with
		$t_1$;$t_2$;$t_3$;$t_4$ referring to the hopping parameters between the highest occupied molecular orbitals. The green circles evidence the dimeric structure. (b-d): 
		Possible structures of the CO phase. Magenta and cyan colored molecules indicate different charge. In (b),
		($P2/c$ space group)  molecules with the same charge form a stripe along $c$ axis, in (c) and (d) ($P2_1/n$ and $P\overline{1}$ groups, respectively) the stripe is along the $b$ axis.}
	\label{fig:4structures}
\end{figure}

\begin{figure}
	\centering
	\includegraphics[width=0.8\linewidth]{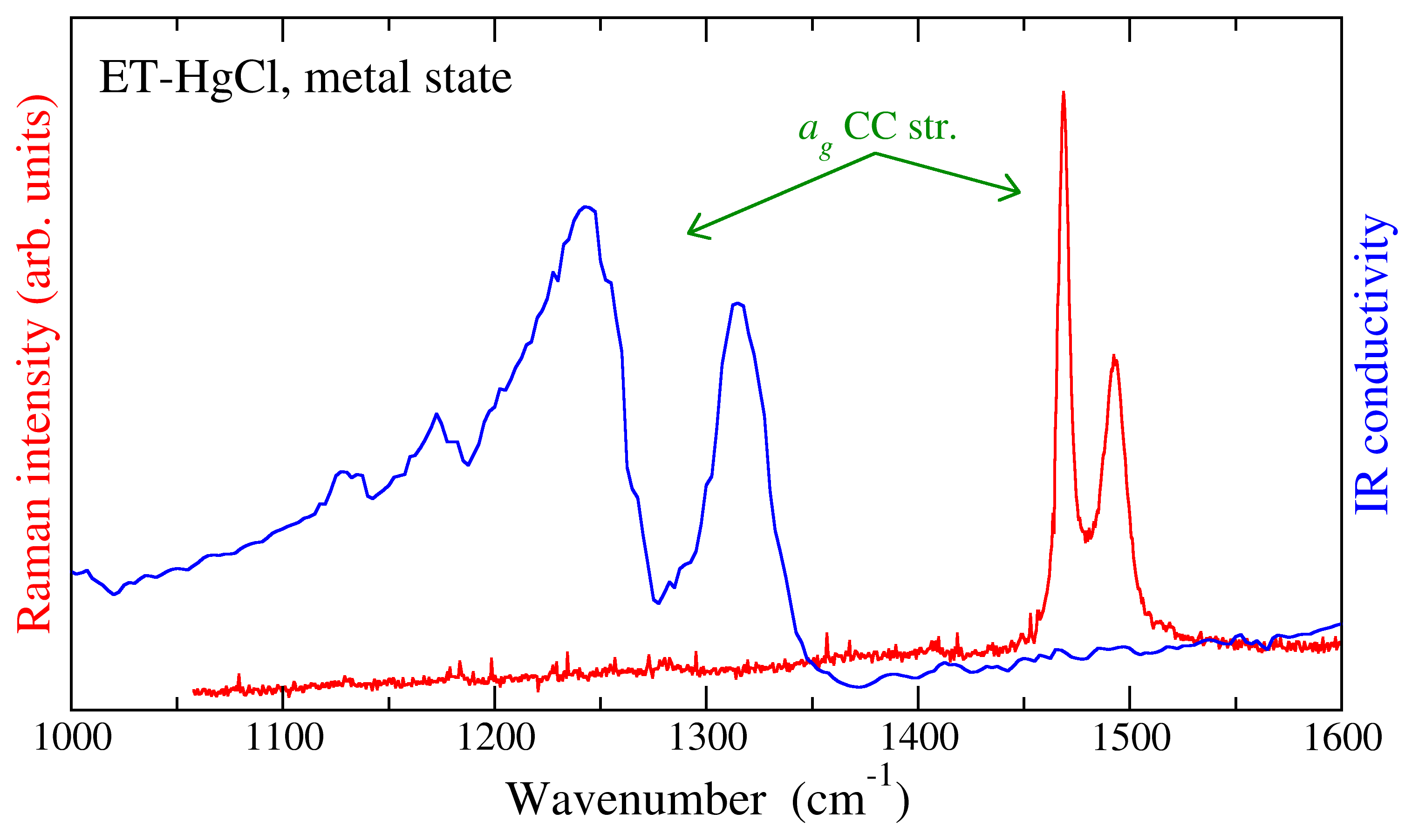}
	\caption{Raman ($cc$ polarization) and IR conductivity spectra ($c$ polarization) of $\kappa$-HgCl at 50 K. The bands associated with i-ph and o-ph $a_g \nu_3$ C=C stretching are evidenced. From Ref. \cite{hassan20}  and \cite{drichko14}, respectively.}
	\label{fig:HgCl50K}
\end{figure}

In the $\kappa$-phase salts the ET molecules are
arranged into 2D layers separated by the counterions.
In each layers the ET molecules are arranged in dimers,
with +1$e$ average charge (0.5$e$ per molecule), disposed
in a chessboard pattern. Figure \ref{fig:HgCl50K}(a)
reports one layer of $\kappa$-HgCl in the metallic, non-CO phase,
space group $C2/c$, viewed along the ET long axis, almost perpendicular to the most developed $bc$ crystal plane.
In the Figure the four nearest-neighbor hopping integrals are
indicated. The intra-dimer one, $t_1$, is estimated 2-3 times larger
than the inter-dimer hopping integrals, justifying
the use of the ``effective-dimer'' model to characterize
the electronic structure \cite{gati18}. Accordingly, the 1D dimer approximation of Figure \ref{fig:4stacks}(b) can be used to interpret
the optical spectra. In fact, the ET layer is 2D, but the optical spectra
are collected along the almost orthogonal $b$ or $c$ axis.
This is clearly illustrated in Figure \ref{fig:HgCl50K}, comparing
the polarized Raman and conductivity spectra of $\kappa$-HgCl
in the metallic (50 K) phase.       

The $\kappa$-HgCl IR conductivity with polarization along the $c$ axis and is characterized by a very intense and broad (more than 100 cm$^{-1}$, with Fano indentation around 1270 cm$^{-1}$) band,
attributed to the o-ph, e-mv coupled mode associated with the
Raman i-ph $a_g \nu_3$ C=C stretching \cite{drichko14,hassan20}. The 
pattern is analogous to that of RbTCNQ(II) LT phase (Figure \ref{fig:RbTCNQrir}b). In the case of $\kappa$-HgCl the IR band is so
largely downshifted, intense and broad because the $a_g \nu_3$ phonon is 
strongly coupled to the very close CT transition, located around 2300 cm$^{-1}$. Analogous feature is observed in the spectra polarized along the $b$ axis.

Raman \cite{hassan20} and  IR \cite{drichko14} spectra showed that     
$\kappa$-HgCl displays a CO transition at 30 K, with  
ET molecules having different charges, 0.6$e$ and 0.4$e$.
Apparently, the space group does not change across the CO
transition \cite{drichko14}, but
assessing subtle differences in the ET molecules, with
disordered ethylene groups, may turn to be difficult.
Excluding cell doubling, as later demonstrated by low-frequency Raman \cite{hassan20}, two possible symmetry descent group were considered \cite{drichko14}, both implying intra-dimer charge separation, as illustrated in Figure \ref{fig:4structures} (b) and (c). The (b) $P2/c$ space group was chosen as the most plausible on the basis of the anisotropy observed in the conductivity spectra. Subsequent 
directional-dependent thermal expansion measurements \cite{gati18} 
showed that changes in the lattice parameters were
consistent with the suggested CO pattern.
\begin{figure}
	\centering
	\includegraphics[width=0.9\linewidth]{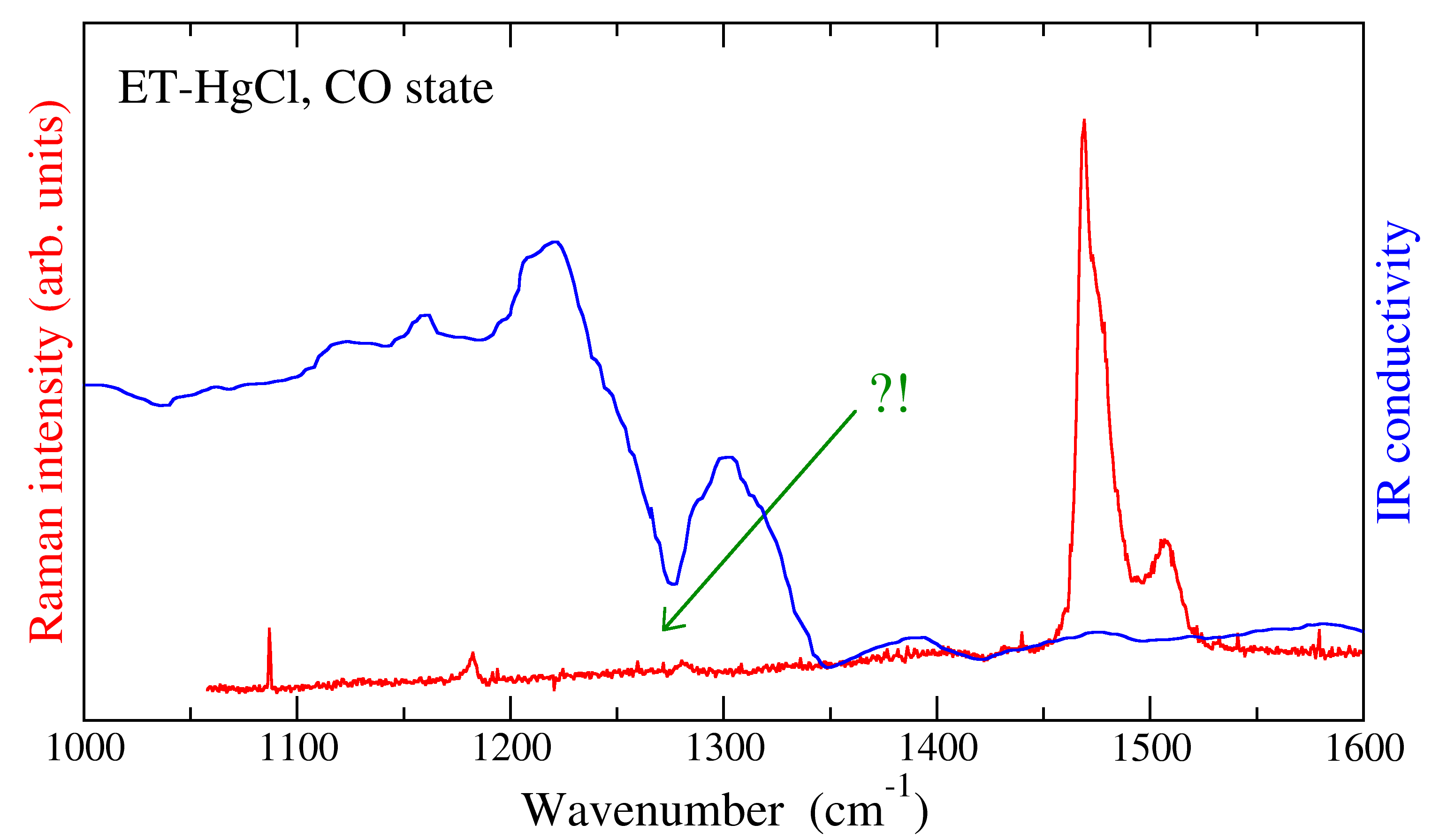}
	\caption{{Raman ($cc$ polarization) and IR conductivity spectra ($c$ polarization) of $\kappa$-HgCl at 20 K. The green arrow underlines the absence of the o-ph $a_g \nu_3$ C=C stretching in Raman. From Ref. \cite{hassan20,girlando24}  and \cite{drichko14}, respectively.}}
	\label{fig:HgCl20K}
\end{figure}

Intra-dimer CO means that the intra-dimer inversion center is
lost, i.e., the packing pattern changes from (b) to (d) in 
Figure \ref{fig:4stacks}: No inversion center is present and all
the phonons become Raman and IR active at the same frequency. As shown
in Figure \ref{fig:DMTTF-CA}d, both spectra are dominated by
the e-mv induced modes, so in the case of ET dimer the
best option is to look for the presence in Raman of the strong o-ph
coupled $a_g \nu_3$ C=C stretching vibration, occurring between 1200 and 1300 cm$^{-1}$. This spectral region is also free from interference
by other Raman bands. Figure \ref{fig:HgCl20K}, reporting
the Raman and IR spectra of $\kappa$-HgCl at 20 K,
below the phase transition, shows no sign of the
expected Raman band associated with o-ph $a_g \nu_3$ stretching. This fact implies that the intra-dimer inversion center is retained,
and that the packing pattern is that of Figure \ref{fig:4structures}(d), space group $P\overline{1}$ \cite{girlando24}. In other words, the absence of the o-ph $a_g \nu_3$ mode in the Raman spectra implies that the
charge disproportionation is inter-dimer, not intra-dimer. 

The possibility of inter-dimer CO pattern in $\kappa$-phase
salts of ET and analogues has been rarely investigated
theoretically \cite{kaneko17,priya26}, also because it was
believed not present experimentally in known materials \cite{lang25}.
The change of perspective suggested above on the basis
of a simple spectroscopic selection rule clearly requires further
examination.

Another ET salt receiving overwhelming attention is the
so-called spin-liquid candidate, $\kappa$-(ET)$_2$Cu$_2$(CN)$_3$ ($\kappa$-CuCN). In this salt, a Mott insulator close to
the superconducting state, the layers of (ET)$_2$ dimers
are arranged in an almost ideal triangular lattice, suggesting
that geometrical frustration suppresses magnetic order 
state. However, the nature
of $\kappa$-CuCN ground state is matter of debate \cite{ilakovac17,pustogow22}.
A recent accurate analysis of the $a_g \nu_2$ C=C stretching Raman bandshape 
has suggested the presence of fluctuating charge disproportionation of amplitude $\simeq 0.05e$ \cite{liebman24}. Since the
reported $\kappa$-CuCN crystal structure ($P2_1$) is similar 
to the $C2/c$ of $\kappa$-HgCl (the main difference resides in the
unit cell, containing one layers of ET rather than two),
it was quite natural to assume intra-dimer CO fluctuations.
Unfortunately, in the literature there is no
relevant analysis or figure concerning the Raman spectral region (1200-1300 cm${-1}$) where the o-ph $a_g \nu_3$ C=C stretching should occur in case of loss of intra-dimer inversion center  \cite{sedlmeier12,yakushi15,liebman24}.Therefore intra-dimer charge fluctuation remains an hypothesis.

On the other hand, recent accurate synchrotron
structural analysis \cite{foury-leylekian18}, followed
by x-ray absorption and inelastic scattering \cite{bayo25}, have shown that the actual $\kappa$-CuCN structure is triclinic, $P\overline{1}$, with two non-equivalent centro-symmetric dimers in the unit cell.
In other words, the structural pattern is the same as that hypothesized
here for $\kappa$-HgCl, Figure \ref{fig:4structures}(d). The
amplitude of the charge disproportionation between the dimers, as estimated from the structural analysis, is $0.06e$,
practically identical to that independently estimated
from Raman. Although the contemporary  loss of intra-dimer inversion center cannot be excluded from the synchrotron data investigation, 
in the lack of contrasting evidence the Occam
razor principle leads to conclude that the charge fluctuation and disproportionation in $\kappa$-CuCN is inter-dimer.

In the $\kappa$-ET family the occurrence of 
CO phenomena are rare, or perhaps difficult to be recognized unless
the magnitude of disproportionation is large \cite{sedlmeier12}. Indeed,
all the following additional examples, as far as I have been able to find in the literature, imply a charge
disproportionation of $\simeq 1e$, that is, the CO state
is made up of fully charged and neutral molecules:
$\kappa$-(ET)$_4$[PtC1$_6$]$\cdot$C$_6$H$_5$CN \cite{galimzyanov89}; triclinic $\kappa$-(ET)$_4$[M(CN)$_6$][N(C$_2$H$_5$)$_4$]$\cdot$3H$_2$O,
M=Fe$^{\mathrm{III}}$, Co$^{\mathrm{III}}$ \cite{swietlik01,swietlik04}; monoclinic 
$\kappa$-(ET)$_4$ [M(CN)$_6$][N(C$_2$H$_5$)$_4]\cdot$2H$_2$O,
 M=Fe$^{\mathrm{III}}$,Co$^{\mathrm{III}}$ \cite{swietlik06,ota07}; 
$\kappa$-(ET)$_2$K$^+$(18-crown-6)[Co$^{\mathrm{II}}$(NCS)$_4$]$\cdot$H$_2$O \cite{bardin23}. Given the strong molecular conformational
difference between neutral and fully charged ET, the performed
x-ray structural analysis easily shows that in {\it all} these compounds
the CO pattern is inter-dimer, with intra-dimer inversion center retained. Spectroscopic measurements have confirmed
the magnitude of disproportionation, and of course do not show trace of the
o-ph $a_g \nu_3$ vibration in the Raman spectra \cite{swietlik01,swietlik04}.
 
An old saying, popular among spectroscopists, went: ``Just because you don't see a band, doesn't mean it isn't there.'' On the other hand,
in the few case of CO in $\kappa$ phase, I have been unable to find a positive
example, namely a case in which the o-ph $a_g \nu_3$ Raman band is detected
in the presence of static/fluctuating CO.

\begin{figure}
	\centering
	\includegraphics[width=1.0\linewidth]{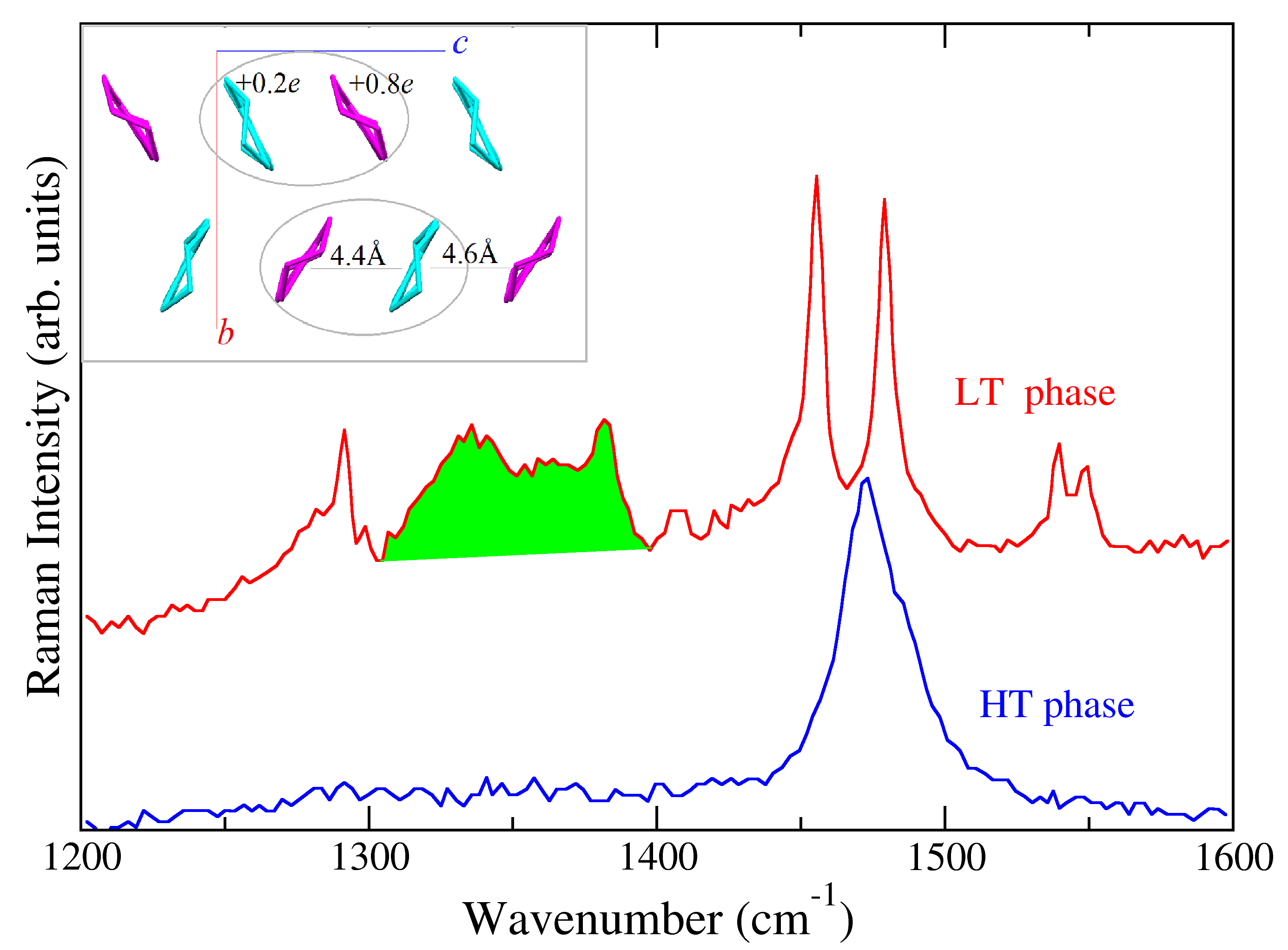}
	\caption{Room temperature and 50 K Raman spectra of $\theta$-(ET)$_2$RbZn(SCN)$_4$. The shaded green area stress the
	appearance of the o-ph $a_g \nu_3$ mode in the CO phase. Adapted from
	Ref.\cite{takahashi06}. The inset is a cartoon of the $P2_12_12_1$ phase, evidencing the intra-dimer CO pattern.}
	\label{fig:et-rbzn}
\end{figure}

 The CO transition
is more common in the so-called $\theta$-phase, whose packing is characterized
by two parallelel stacks with tilted ET molecules
arranged in brick-wall-like pattern. A clean, well characterized CO example  can be found in $\theta$-(ET)$_2$RbZn(SCN)$_4$, that at room
temperature crystallizes in the $I222$ space grups, two stacks in the unit cell, with equal distance among the molecules, i.e., type (a) stack
in Figure \ref{fig:4stacks} \cite{watanabe04}. At 90 K 
$\theta$-(ET)$_2$RbZn(SCN)$_4$ undergoes a phase transition to the
$P2_12_12_1$ space group, implying at the same time stack dimerization
{\it and} intra-dimer CO (+0.2$e$ and +0.8$e$, see inset of Figure
\ref{fig:et-rbzn}). The structure is more complex than that of the $\kappa$-phase,
but what it is important here is that in the low-temperature phase all the inversion centers along the stacks are lost, Figure \ref{fig:4stacks}(d), so the IR active, e-mv coupled o-ph $a_g \nu_3$ C=C stretching should appear also in Raman. Figure \ref{fig:et-rbzn} shows that indeed this is the case:
a broad Raman band, not present at room temperature, appears in the
expected spectral region in the low
temperature, intra-dimer CO phase \cite{takahashi06}.

\textit{Conclusions}--In the field of low-dimensional CT crystals it is well known that the Peierls distortion of a uniform chain of molecules
(loss of inversion center on the sites) gives rise to the IR ``activation'' of the o-ph $a_g$ modes, e-mv coupled with the CT electron. Far less familiar is that the same o-ph $a_g$ modes become ``activated''
in Raman when a intra-dimer CO transition occurs in a dimerized chain of identical molecules (loss of inversion center between the sites.) Perhaps for this reason the Raman spectral region where the ET most strongly e-mv coupled phonon, the $a_g \nu_3$, should occur in case of inter-dimer CO, did not receive the deserved attention. As a matter of fact, the presence of a
Raman band associated to this mode has never been reported as a proof
of intra-dimer dispropotionation in the the CO phase of $\kappa$-ET
salts and analogues. Recording a Raman spectrum is sometimes tricky, as
self-absorption of emitted radiation, and/or resonance effects may obscure
the results, but a deeper investigation is certainly worthwhile to
provide a model independent proof of intra-dimer CO.

In the lack of such a proof, and considering that whenever
adequately accurate structural analysis are available, all
indicate inter-dimer CO, I suggest that this is indeed the
prevailing, if not unique, CO pattern in $\kappa$-phase
salts. This suggestion implies a complete change of paradigm,
and challenges theory to explain the multivariate and unique
phenomena exhibited by ET (and molecular analogues) 
low-dimensional CT crystals \cite{clay19,dressel20,lang25}.

\begin{acknowledgments}
\textit{Acknowledgments}---I thank Matteo Masino 
(Parma University) and Natalia Drichko (Hopkins University at Baltimore) for discussions and for providing the original Raman data of RbTCNQ(II) and ET-HgCl. I am also grateful to Martin Dressel (Stuttgart University)
for encouragement and correspondence.
\end{acknowledgments}

\begin{thebibliography}{39}%
\makeatletter
\providecommand \@ifxundefined [1]{%
 \@ifx{#1\undefined}
}%
\providecommand \@ifnum [1]{%
 \ifnum #1\expandafter \@firstoftwo
 \else \expandafter \@secondoftwo
 \fi
}%
\providecommand \@ifx [1]{%
 \ifx #1\expandafter \@firstoftwo
 \else \expandafter \@secondoftwo
 \fi
}%
\providecommand \natexlab [1]{#1}%
\providecommand \enquote  [1]{``#1''}%
\providecommand \bibnamefont  [1]{#1}%
\providecommand \bibfnamefont [1]{#1}%
\providecommand \citenamefont [1]{#1}%
\providecommand \href@noop [0]{\@secondoftwo}%
\providecommand \href [0]{\begingroup \@sanitize@url \@href}%
\providecommand \@href[1]{\@@startlink{#1}\@@href}%
\providecommand \@@href[1]{\endgroup#1\@@endlink}%
\providecommand \@sanitize@url [0]{\catcode `\\12\catcode `\$12\catcode
  `\&12\catcode `\#12\catcode `\^12\catcode `\_12\catcode `\%12\relax}%
\providecommand \@@startlink[1]{}%
\providecommand \@@endlink[0]{}%
\providecommand \url  [0]{\begingroup\@sanitize@url \@url }%
\providecommand \@url [1]{\endgroup\@href {#1}{\urlprefix }}%
\providecommand \urlprefix  [0]{URL }%
\providecommand \Eprint [0]{\href }%
\providecommand \doibase [0]{https://doi.org/}%
\providecommand \selectlanguage [0]{\@gobble}%
\providecommand \bibinfo  [0]{\@secondoftwo}%
\providecommand \bibfield  [0]{\@secondoftwo}%
\providecommand \translation [1]{[#1]}%
\providecommand \BibitemOpen [0]{}%
\providecommand \bibitemStop [0]{}%
\providecommand \bibitemNoStop [0]{.\EOS\space}%
\providecommand \EOS [0]{\spacefactor3000\relax}%
\providecommand \BibitemShut  [1]{\csname bibitem#1\endcsname}%
\let\auto@bib@innerbib\@empty
\bibitem [{\citenamefont {Dressel}\ and\ \citenamefont
  {Tomi{\'{c}}}(2020)}]{dressel20}%
  \BibitemOpen
  \bibfield  {author} {\bibinfo {author} {\bibfnamefont {M.}~\bibnamefont
  {Dressel}}\ and\ \bibinfo {author} {\bibfnamefont {S.}~\bibnamefont
  {Tomi{\'{c}}}},\ }\bibfield  {title} {\bibinfo {title} {Molecular quantum
  materials: electronic phases and charge dynamics in two-dimensional organic
  solids},\ }\href {https://doi.org/10.1080/00018732.2020.1837833} {\bibfield
  {journal} {\bibinfo  {journal} {Adv. Phys.}\ }\textbf {\bibinfo {volume}
  {69}},\ \bibinfo {pages} {1} (\bibinfo {year} {2020})}\BibitemShut {NoStop}%
\bibitem [{\citenamefont {Clay}\ and\ \citenamefont {Mazumdar}(2019)}]{clay19}%
  \BibitemOpen
  \bibfield  {author} {\bibinfo {author} {\bibfnamefont {R.}~\bibnamefont
  {Clay}}\ and\ \bibinfo {author} {\bibfnamefont {S.}~\bibnamefont
  {Mazumdar}},\ }\bibfield  {title} {\bibinfo {title} {From charge- and
  spin-ordering to superconductivity in the organic charge-transfer solids},\
  }\href {https://doi.org/10.1016/j.physrep.2018.10.006} {\bibfield  {journal}
  {\bibinfo  {journal} {Phys. Rep.}\ }\textbf {\bibinfo {volume} {788}},\
  \bibinfo {pages} {1} (\bibinfo {year} {2019})}\BibitemShut {NoStop}%
\bibitem [{\citenamefont {Lang}\ \emph {et~al.}(2025)\citenamefont {Lang},
  \citenamefont {Lunkenheimer}, \citenamefont {Ganter}, \citenamefont
  {Winter},\ and\ \citenamefont {Müller}}]{lang25}%
  \BibitemOpen
  \bibfield  {author} {\bibinfo {author} {\bibfnamefont {M.}~\bibnamefont
  {Lang}}, \bibinfo {author} {\bibfnamefont {P.}~\bibnamefont {Lunkenheimer}},
  \bibinfo {author} {\bibfnamefont {O.}~\bibnamefont {Ganter}}, \bibinfo
  {author} {\bibfnamefont {S.}~\bibnamefont {Winter}},\ and\ \bibinfo {author}
  {\bibfnamefont {J.}~\bibnamefont {Müller}},\ }\bibfield  {title} {\bibinfo
  {title} {Ferroelectric and multiferroic properties of quasi-2d organic
  charge-transfer salts: A review},\ }\href
  {https://doi.org/10.1007/s11664-025-11756-3} {\bibfield  {journal} {\bibinfo
  {journal} {J. Electron. Mater.}\ }\textbf {\bibinfo {volume} {54}},\ \bibinfo
  {pages} {5087} (\bibinfo {year} {2025})}\BibitemShut {NoStop}%
\bibitem [{\citenamefont {Canossa}\ \emph {et~al.}(2021)\citenamefont
  {Canossa}, \citenamefont {Ferrari}, \citenamefont {Sippel}, \citenamefont
  {Fischer}, \citenamefont {Pfattner}, \citenamefont {Frison}, \citenamefont
  {Masino}, \citenamefont {Mas-Torrent}, \citenamefont {Lunkenheimer},
  \citenamefont {Rovira},\ and\ \citenamefont {Girlando}}]{canossa21}%
  \BibitemOpen
  \bibfield  {author} {\bibinfo {author} {\bibfnamefont {S.}~\bibnamefont
  {Canossa}}, \bibinfo {author} {\bibfnamefont {E.}~\bibnamefont {Ferrari}},
  \bibinfo {author} {\bibfnamefont {P.}~\bibnamefont {Sippel}}, \bibinfo
  {author} {\bibfnamefont {J.~K.~H.}\ \bibnamefont {Fischer}}, \bibinfo
  {author} {\bibfnamefont {R.}~\bibnamefont {Pfattner}}, \bibinfo {author}
  {\bibfnamefont {R.}~\bibnamefont {Frison}}, \bibinfo {author} {\bibfnamefont
  {M.}~\bibnamefont {Masino}}, \bibinfo {author} {\bibfnamefont
  {M.}~\bibnamefont {Mas-Torrent}}, \bibinfo {author} {\bibfnamefont
  {P.}~\bibnamefont {Lunkenheimer}}, \bibinfo {author} {\bibfnamefont
  {C.}~\bibnamefont {Rovira}},\ and\ \bibinfo {author} {\bibfnamefont
  {A.}~\bibnamefont {Girlando}},\ }\bibfield  {title} {\bibinfo {title}
  {Tetramethylbenzidine{\textendash}{TetrafluoroTCNQ}
  ({TMB}{\textendash}{TCNQF}\textsubscript{4}): A narrow-gap semiconducting
  salt with room-temperature relaxor ferroelectric behavior},\ }\href
  {https://doi.org/10.1021/acs.jpcc.1c07131} {\bibfield  {journal} {\bibinfo
  {journal} {J. Phys. Chem. C}\ }\textbf {\bibinfo {volume} {125}},\ \bibinfo
  {pages} {25816} (\bibinfo {year} {2021})}\BibitemShut {NoStop}%
\bibitem [{\citenamefont {Salzillo}\ \emph {et~al.}(2021)\citenamefont
  {Salzillo}, \citenamefont {Giunchi}, \citenamefont {Pandolfi}, \citenamefont
  {Brillante},\ and\ \citenamefont {Venuti}}]{salzillo21}%
  \BibitemOpen
  \bibfield  {author} {\bibinfo {author} {\bibfnamefont {T.}~\bibnamefont
  {Salzillo}}, \bibinfo {author} {\bibfnamefont {A.}~\bibnamefont {Giunchi}},
  \bibinfo {author} {\bibfnamefont {L.}~\bibnamefont {Pandolfi}}, \bibinfo
  {author} {\bibfnamefont {A.}~\bibnamefont {Brillante}},\ and\ \bibinfo
  {author} {\bibfnamefont {E.}~\bibnamefont {Venuti}},\ }\bibfield  {title}
  {\bibinfo {title} {Bulk and surface-mediated polymorphs of bio-inspired dyes
  organic semiconductors: The role of lattice phonons in their investigation},\
  }\href {https://doi.org/10.1002/ijch.202100067} {\bibfield  {journal}
  {\bibinfo  {journal} {Isr. J. Chem.}\ }\textbf {\bibinfo {volume} {61}},\
  \bibinfo {pages} {650} (\bibinfo {year} {2021})}\BibitemShut {NoStop}%
\bibitem [{\citenamefont {Brillante}\ \emph {et~al.}(2008)\citenamefont
  {Brillante}, \citenamefont {Bilotti}, \citenamefont {Valle}, \citenamefont
  {Venuti},\ and\ \citenamefont {Girlando}}]{brillante08}%
  \BibitemOpen
  \bibfield  {author} {\bibinfo {author} {\bibfnamefont {A.}~\bibnamefont
  {Brillante}}, \bibinfo {author} {\bibfnamefont {I.}~\bibnamefont {Bilotti}},
  \bibinfo {author} {\bibfnamefont {R.~G.~D.}\ \bibnamefont {Valle}}, \bibinfo
  {author} {\bibfnamefont {E.}~\bibnamefont {Venuti}},\ and\ \bibinfo {author}
  {\bibfnamefont {A.}~\bibnamefont {Girlando}},\ }\bibfield  {title} {\bibinfo
  {title} {Probing polymorphs of organic semiconductors by lattice phonon raman
  microscopy},\ }\href {https://doi.org/10.1039/b804317e} {\bibfield  {journal}
  {\bibinfo  {journal} {{CrystEngComm}}\ }\textbf {\bibinfo {volume} {10}},\
  \bibinfo {pages} {937} (\bibinfo {year} {2008})}\BibitemShut {NoStop}%
\bibitem [{\citenamefont {Girlando}\ \emph {et~al.}(2016)\citenamefont
  {Girlando}, \citenamefont {Masino}, \citenamefont {Brillante}, \citenamefont
  {Toccoli},\ and\ \citenamefont {Iannotta}}]{girlando16}%
  \BibitemOpen
  \bibfield  {author} {\bibinfo {author} {\bibfnamefont {A.}~\bibnamefont
  {Girlando}}, \bibinfo {author} {\bibfnamefont {M.}~\bibnamefont {Masino}},
  \bibinfo {author} {\bibfnamefont {A.}~\bibnamefont {Brillante}}, \bibinfo
  {author} {\bibfnamefont {T.}~\bibnamefont {Toccoli}},\ and\ \bibinfo {author}
  {\bibfnamefont {S.}~\bibnamefont {Iannotta}},\ }\bibfield  {title} {\bibinfo
  {title} {Raman identification of polymorphs in pentacene films},\ }\href
  {https://doi.org/10.3390/cryst6040041} {\bibfield  {journal} {\bibinfo
  {journal} {Crystals}\ }\textbf {\bibinfo {volume} {6}},\ \bibinfo {pages}
  {41} (\bibinfo {year} {2016})}\BibitemShut {NoStop}%
\bibitem [{\citenamefont {Turrell}(1972)}]{turrell72}%
  \BibitemOpen
  \bibfield  {author} {\bibinfo {author} {\bibfnamefont {G.}~\bibnamefont
  {Turrell}},\ }\href@noop {} {\emph {\bibinfo {title} {\textit{Infrared and
  Raman Spectra of Crystals}}}}\ (\bibinfo  {publisher} {Academic Press},\
  \bibinfo {address} {London and New York},\ \bibinfo {year}
  {1972})\BibitemShut {NoStop}%
\bibitem [{\citenamefont {Painelli}\ and\ \citenamefont
  {Girlando}(1986)}]{painelli86}%
  \BibitemOpen
  \bibfield  {author} {\bibinfo {author} {\bibfnamefont {A.}~\bibnamefont
  {Painelli}}\ and\ \bibinfo {author} {\bibfnamefont {A.}~\bibnamefont
  {Girlando}},\ }\bibfield  {title} {\bibinfo {title} {{Electron–molecular
  vibration (e–mv) coupling in charge‐transfer compounds and its
  consequences on the optical spectra: A theoretical framework}},\ }\href
  {https://doi.org/10.1063/1.449926} {\bibfield  {journal} {\bibinfo  {journal}
  {J. Chem. Phys.}\ }\textbf {\bibinfo {volume} {84}},\ \bibinfo {pages} {5655}
  (\bibinfo {year} {1986})}\BibitemShut {NoStop}%
\bibitem [{\citenamefont {Pecile}\ \emph {et~al.}(1989)\citenamefont {Pecile},
  \citenamefont {Painelli},\ and\ \citenamefont {Girlando}}]{pecile89}%
  \BibitemOpen
  \bibfield  {author} {\bibinfo {author} {\bibfnamefont {C.}~\bibnamefont
  {Pecile}}, \bibinfo {author} {\bibfnamefont {A.}~\bibnamefont {Painelli}},\
  and\ \bibinfo {author} {\bibfnamefont {A.}~\bibnamefont {Girlando}},\
  }\bibfield  {title} {\bibinfo {title} {Studies of organic semiconductors for
  40 years—v},\ }\href {https://doi.org/10.1080/00268948908065787} {\bibfield
   {journal} {\bibinfo  {journal} {Mol. Cryst. Liq. Cryst. Incorporating
  Nonlinear Opt.}\ }\textbf {\bibinfo {volume} {171}},\ \bibinfo {pages} {69}
  (\bibinfo {year} {1989})}\BibitemShut {NoStop}%
\bibitem [{\citenamefont {Bozio}\ and\ \citenamefont {Pecile}(1991)}]{bozio91}%
  \BibitemOpen
  \bibfield  {author} {\bibinfo {author} {\bibfnamefont {R.}~\bibnamefont
  {Bozio}}\ and\ \bibinfo {author} {\bibfnamefont {C.}~\bibnamefont {Pecile}},\
  }\href@noop {} {\emph {\bibinfo {title} {Charge transfer crystals and
  molecular conductors}}},\ edited by\ \bibinfo {editor} {\bibfnamefont
  {R.~J.~H.}\ \bibnamefont {Clark}}\ and\ \bibinfo {editor} {\bibfnamefont
  {R.~E.}\ \bibnamefont {Hester}},\ \bibinfo {series} {Spectroscopy of Advanced
  Materials}, Vol.~\bibinfo {volume} {19}\ (\bibinfo  {publisher} {John
  Wiley},\ \bibinfo {year} {1991})\ Chap.~\bibinfo {chapter} {1}, pp.\ \bibinfo
  {pages} {1--86}\BibitemShut {NoStop}%
\bibitem [{\citenamefont {Tomi{\'{c}}}\ and\ \citenamefont
  {Dressel}(2015)}]{tomic15}%
  \BibitemOpen
  \bibfield  {author} {\bibinfo {author} {\bibfnamefont {S.}~\bibnamefont
  {Tomi{\'{c}}}}\ and\ \bibinfo {author} {\bibfnamefont {M.}~\bibnamefont
  {Dressel}},\ }\bibfield  {title} {\bibinfo {title} {Ferroelectricity in
  molecular solids: a review of electrodynamic properties},\ }\href
  {https://doi.org/10.1088/0034-4885/78/9/096501} {\bibfield  {journal}
  {\bibinfo  {journal} {Rep. Prog. Phys.}\ }\textbf {\bibinfo {volume} {78}},\
  \bibinfo {pages} {096501} (\bibinfo {year} {2015})}\BibitemShut {NoStop}%
\bibitem [{\citenamefont {Bozio}\ and\ \citenamefont {Pecile}(1997)}]{bozio97}%
  \BibitemOpen
  \bibfield  {author} {\bibinfo {author} {\bibfnamefont {R.}~\bibnamefont
  {Bozio}}\ and\ \bibinfo {author} {\bibfnamefont {C.}~\bibnamefont {Pecile}},\
  }\bibfield  {title} {\bibinfo {title} {Phase transitions of (1:1) alkaline
  salts of {TCNQ} as studied by vibronic intensity enhancement in the infrared
  spectra},\ }\href@noop {} {\bibfield  {journal} {\bibinfo  {journal} {J.
  Chem. Phys.}\ }\textbf {\bibinfo {volume} {67}},\ \bibinfo {pages} {3864}
  (\bibinfo {year} {1997})}\BibitemShut {NoStop}%
\bibitem [{\citenamefont {McQueen}\ \emph {et~al.}(2009)\citenamefont
  {McQueen}, \citenamefont {Ho}, \citenamefont {Cahua}, \citenamefont {Cava},
  \citenamefont {Pascal},\ and\ \citenamefont {Soos}}]{mcqueen09}%
  \BibitemOpen
  \bibfield  {author} {\bibinfo {author} {\bibfnamefont {T.}~\bibnamefont
  {McQueen}}, \bibinfo {author} {\bibfnamefont {D.}~\bibnamefont {Ho}},
  \bibinfo {author} {\bibfnamefont {C.~J.}\ \bibnamefont {Cahua}}, \bibinfo
  {author} {\bibfnamefont {R.}~\bibnamefont {Cava}}, \bibinfo {author}
  {\bibfnamefont {R.}~\bibnamefont {Pascal}},\ and\ \bibinfo {author}
  {\bibfnamefont {Z.}~\bibnamefont {Soos}},\ }\bibfield  {title} {\bibinfo
  {title} {Realization of the bond order wave ({BOW}) phase of extended
  {H}ubbard models in {Rb}{TCNQ}({II})},\ }\href
  {https://doi.org/10.1016/j.cplett.2009.04.085} {\bibfield  {journal}
  {\bibinfo  {journal} {Chem. Phys. Lett.}\ }\textbf {\bibinfo {volume}
  {475}},\ \bibinfo {pages} {44} (\bibinfo {year} {2009})}\BibitemShut
  {NoStop}%
\bibitem [{\citenamefont {Ranzieri}\ \emph {et~al.}(2007)\citenamefont
  {Ranzieri}, \citenamefont {Masino}, \citenamefont {Girlando},\ and\
  \citenamefont {Lem{\'{e}}e-Cailleau}}]{ranzieri07a}%
  \BibitemOpen
  \bibfield  {author} {\bibinfo {author} {\bibfnamefont {P.}~\bibnamefont
  {Ranzieri}}, \bibinfo {author} {\bibfnamefont {M.}~\bibnamefont {Masino}},
  \bibinfo {author} {\bibfnamefont {A.}~\bibnamefont {Girlando}},\ and\
  \bibinfo {author} {\bibfnamefont {M.~H.}\ \bibnamefont
  {Lem{\'{e}}e-Cailleau}},\ }\bibfield  {title} {\bibinfo {title}
  {{Temperature-induced valence and structural instability in DMTTF-CA:
  Single-crystal Raman and infrared measurements}},\ }\href
  {https://doi.org/10.1103/PhysRevB.76.134115} {\bibfield  {journal} {\bibinfo
  {journal} {Phys. Rev. B}\ }\textbf {\bibinfo {volume} {76}},\ \bibinfo
  {pages} {1} (\bibinfo {year} {2007})}\BibitemShut {NoStop}%
\bibitem [{\citenamefont {Drichko}\ \emph {et~al.}(2014)\citenamefont
  {Drichko}, \citenamefont {Beyer}, \citenamefont {Rose}, \citenamefont
  {Dressel}, \citenamefont {Schlueter}, \citenamefont {Turunova}, \citenamefont
  {Zhilyaeva},\ and\ \citenamefont {Lyubovskaya}}]{drichko14}%
  \BibitemOpen
  \bibfield  {author} {\bibinfo {author} {\bibfnamefont {N.}~\bibnamefont
  {Drichko}}, \bibinfo {author} {\bibfnamefont {R.}~\bibnamefont {Beyer}},
  \bibinfo {author} {\bibfnamefont {E.}~\bibnamefont {Rose}}, \bibinfo {author}
  {\bibfnamefont {M.}~\bibnamefont {Dressel}}, \bibinfo {author} {\bibfnamefont
  {J.~A.}\ \bibnamefont {Schlueter}}, \bibinfo {author} {\bibfnamefont {S.~A.}\
  \bibnamefont {Turunova}}, \bibinfo {author} {\bibfnamefont {E.~I.}\
  \bibnamefont {Zhilyaeva}},\ and\ \bibinfo {author} {\bibfnamefont {R.~N.}\
  \bibnamefont {Lyubovskaya}},\ }\bibfield  {title} {\bibinfo {title} {Metallic
  state and charge-order metal-insulator transition in the
  quasi-two-dimensional conductor
  $\kappa$-{(BEDT-TTF)}\textsubscript{2}{Hg}({SCN})\textsubscript{2}{Cl}},\
  }\href {https://doi.org/10.1103/physrevb.89.075133} {\bibfield  {journal}
  {\bibinfo  {journal} {Phys. Rev. B}\ }\textbf {\bibinfo {volume} {89}},\
  \bibinfo {pages} {075133} (\bibinfo {year} {2014})}\BibitemShut {NoStop}%
\bibitem [{\citenamefont {Gati}\ \emph {et~al.}(2018)\citenamefont {Gati},
  \citenamefont {Fischer}, \citenamefont {Lunkenheimer}, \citenamefont
  {Zielke}, \citenamefont {K\"ohler}, \citenamefont {Kolb}, \citenamefont {von
  Nidda}, \citenamefont {Winter}, \citenamefont {Schubert}, \citenamefont
  {Schlueter}, \citenamefont {Jeschke}, \citenamefont {Valent{\'{\i}}},\ and\
  \citenamefont {Lang}}]{gati18}%
  \BibitemOpen
  \bibfield  {author} {\bibinfo {author} {\bibfnamefont {E.}~\bibnamefont
  {Gati}}, \bibinfo {author} {\bibfnamefont {J.~K.~H.}\ \bibnamefont
  {Fischer}}, \bibinfo {author} {\bibfnamefont {P.}~\bibnamefont
  {Lunkenheimer}}, \bibinfo {author} {\bibfnamefont {D.}~\bibnamefont
  {Zielke}}, \bibinfo {author} {\bibfnamefont {S.}~\bibnamefont {K\"ohler}},
  \bibinfo {author} {\bibfnamefont {F.}~\bibnamefont {Kolb}}, \bibinfo {author}
  {\bibfnamefont {H.-A.~K.}\ \bibnamefont {von Nidda}}, \bibinfo {author}
  {\bibfnamefont {S.~M.}\ \bibnamefont {Winter}}, \bibinfo {author}
  {\bibfnamefont {H.}~\bibnamefont {Schubert}}, \bibinfo {author}
  {\bibfnamefont {J.~A.}\ \bibnamefont {Schlueter}}, \bibinfo {author}
  {\bibfnamefont {H.~O.}\ \bibnamefont {Jeschke}}, \bibinfo {author}
  {\bibfnamefont {R.}~\bibnamefont {Valent{\'{\i}}}},\ and\ \bibinfo {author}
  {\bibfnamefont {M.}~\bibnamefont {Lang}},\ }\bibfield  {title} {\bibinfo
  {title} {{Evidence for Electronically Driven Ferroelectricity in a Strongly
  Correlated Dimerized BEDT-TTF Molecular Conductor}},\ }\href
  {https://doi.org/10.1103/physrevlett.120.247601} {\bibfield  {journal}
  {\bibinfo  {journal} {Phys. Rev. Lett.}\ }\textbf {\bibinfo {volume} {120}},\
  \bibinfo {pages} {247601} (\bibinfo {year} {2018})}\BibitemShut {NoStop}%
\bibitem [{\citenamefont {Thomas}\ \emph {et~al.}(2019)\citenamefont {Thomas},
  \citenamefont {Hartmann}, \citenamefont {Lunkenheimer}, \citenamefont
  {Schubert}, \citenamefont {Schlueter},\ and\ \citenamefont
  {M\"uller}}]{thomas19}%
  \BibitemOpen
  \bibfield  {author} {\bibinfo {author} {\bibfnamefont {T.}~\bibnamefont
  {Thomas}}, \bibinfo {author} {\bibfnamefont {B.}~\bibnamefont {Hartmann}},
  \bibinfo {author} {\bibfnamefont {P.}~\bibnamefont {Lunkenheimer}}, \bibinfo
  {author} {\bibfnamefont {H.}~\bibnamefont {Schubert}}, \bibinfo {author}
  {\bibfnamefont {J.~A.}\ \bibnamefont {Schlueter}},\ and\ \bibinfo {author}
  {\bibfnamefont {J.}~\bibnamefont {M\"uller}},\ }\bibfield  {title} {\bibinfo
  {title} {Low-frequency charge carrier dynamics in ferroelectric
  $\kappa$-({BEDT‐TTF})\textsubscript{2}{X} - a comparative study of {X} =
  {Cu[N(CN)}\textsubscript{2}]{Cl} and {X} = {Hg(SCN)}\textsubscript{2}{Cl}},\
  }\href {https://doi.org/10.1002/pssb.201800746} {\bibfield  {journal}
  {\bibinfo  {journal} {Phys. Status Solidi B}\ }\textbf {\bibinfo {volume}
  {256}},\ \bibinfo {pages} {1800746} (\bibinfo {year} {2019})}\BibitemShut
  {NoStop}%
\bibitem [{\citenamefont {Hassan}\ \emph {et~al.}(2020)\citenamefont {Hassan},
  \citenamefont {Thirunavukkuarasu}, \citenamefont {Lu}, \citenamefont
  {Smirnov}, \citenamefont {Zhilyaeva}, \citenamefont {Torunova}, \citenamefont
  {Lyubovskaya},\ and\ \citenamefont {Drichko}}]{hassan20}%
  \BibitemOpen
  \bibfield  {author} {\bibinfo {author} {\bibfnamefont {N.~M.}\ \bibnamefont
  {Hassan}}, \bibinfo {author} {\bibfnamefont {K.}~\bibnamefont
  {Thirunavukkuarasu}}, \bibinfo {author} {\bibfnamefont {Z.}~\bibnamefont
  {Lu}}, \bibinfo {author} {\bibfnamefont {D.}~\bibnamefont {Smirnov}},
  \bibinfo {author} {\bibfnamefont {E.~I.}\ \bibnamefont {Zhilyaeva}}, \bibinfo
  {author} {\bibfnamefont {S.}~\bibnamefont {Torunova}}, \bibinfo {author}
  {\bibfnamefont {R.~N.}\ \bibnamefont {Lyubovskaya}},\ and\ \bibinfo {author}
  {\bibfnamefont {N.}~\bibnamefont {Drichko}},\ }\bibfield  {title} {\bibinfo
  {title} {Melting of charge order in the low-temperature state of an
  electronic ferroelectric-like system},\ }\href
  {https://doi.org/10.1038/s41535-020-0217-5} {\bibfield  {journal} {\bibinfo
  {journal} {npj Quantum Mater.}\ }\textbf {\bibinfo {volume} {5}},\ \bibinfo
  {pages} {15} (\bibinfo {year} {2020})}\BibitemShut {NoStop}%
\bibitem [{\citenamefont {Liebman}\ \emph {et~al.}(2026)\citenamefont
  {Liebman}, \citenamefont {Torunova}, \citenamefont {Schlueter}, \citenamefont
  {Zhilyaeva},\ and\ \citenamefont {Drichko}}]{liebman26}%
  \BibitemOpen
  \bibfield  {author} {\bibinfo {author} {\bibfnamefont {J.}~\bibnamefont
  {Liebman}}, \bibinfo {author} {\bibfnamefont {S.}~\bibnamefont {Torunova}},
  \bibinfo {author} {\bibfnamefont {J.~A.}\ \bibnamefont {Schlueter}}, \bibinfo
  {author} {\bibfnamefont {E.}~\bibnamefont {Zhilyaeva}},\ and\ \bibinfo
  {author} {\bibfnamefont {N.}~\bibnamefont {Drichko}},\ }\bibfield  {title}
  {\bibinfo {title} {Tuning charge order in mott insulators
  $\kappa$-({BEDT-TTF})\textsubscript{2}{Hg}({SCN})\textsubscript{2}{X} ({X =
  Br,Cl}) via uniaxial strain},\ }\href {https://doi.org/10.1103/qzr7-dd5f}
  {\bibfield  {journal} {\bibinfo  {journal} {Phys. Rev. B}\ }\textbf {\bibinfo
  {volume} {113}},\ \bibinfo {pages} {205141} (\bibinfo {year}
  {2026})}\BibitemShut {NoStop}%
\bibitem [{\citenamefont {Urai}\ \emph {et~al.}(2026)\citenamefont {Urai},
  \citenamefont {Miyagawa}, \citenamefont {Torunova}, \citenamefont {Drichko},
  \citenamefont {Zhilyaeva},\ and\ \citenamefont {Kanoda}}]{urai26}%
  \BibitemOpen
  \bibfield  {author} {\bibinfo {author} {\bibfnamefont {M.}~\bibnamefont
  {Urai}}, \bibinfo {author} {\bibfnamefont {K.}~\bibnamefont {Miyagawa}},
  \bibinfo {author} {\bibfnamefont {S.~A.}\ \bibnamefont {Torunova}}, \bibinfo
  {author} {\bibfnamefont {N.}~\bibnamefont {Drichko}}, \bibinfo {author}
  {\bibfnamefont {E.~I.}\ \bibnamefont {Zhilyaeva}},\ and\ \bibinfo {author}
  {\bibfnamefont {K.}~\bibnamefont {Kanoda}},\ }\bibfield  {title} {\bibinfo
  {title} {$^{13}${C NMR} observation of a nonmagnetic charge-ordered state in
  the organic conductor
  $\kappa$-({ET})\textsubscript{2}{Hg}({SCN})\textsubscript{2}{Cl}},\ }\href
  {https://doi.org/10.1103/k13b-4ch4} {\bibfield  {journal} {\bibinfo
  {journal} {Phys. Rev. B}\ }\textbf {\bibinfo {volume} {113}},\ \bibinfo
  {pages} {094453} (\bibinfo {year} {2026})}\BibitemShut {NoStop}%
\bibitem [{\citenamefont {Girlando}(2024)}]{girlando24}%
  \BibitemOpen
  \bibfield  {author} {\bibinfo {author} {\bibfnamefont {A.}~\bibnamefont
  {Girlando}},\ }\bibfield  {title} {\bibinfo {title} {Raman signatures of the
  strong intramolecular and intermolecular charge oscillations in
  bis(ethylenedithio)-tetrathiafulvalene ({BEDT-TTF}) $\kappa$-phase salts},\
  }\href {https://doi.org/10.1103/physrevb.110.035101} {\bibfield  {journal}
  {\bibinfo  {journal} {Phys. Rev. B}\ }\textbf {\bibinfo {volume} {110}},\
  \bibinfo {pages} {035101} (\bibinfo {year} {2024})}\BibitemShut {NoStop}%
\bibitem [{\citenamefont {Kaneko}\ \emph {et~al.}(2017)\citenamefont {Kaneko},
  \citenamefont {Tocchio}, \citenamefont {Valent{\'{\i}}},\ and\ \citenamefont
  {Becca}}]{kaneko17}%
  \BibitemOpen
  \bibfield  {author} {\bibinfo {author} {\bibfnamefont {R.}~\bibnamefont
  {Kaneko}}, \bibinfo {author} {\bibfnamefont {L.~F.}\ \bibnamefont {Tocchio}},
  \bibinfo {author} {\bibfnamefont {R.}~\bibnamefont {Valent{\'{\i}}}},\ and\
  \bibinfo {author} {\bibfnamefont {F.}~\bibnamefont {Becca}},\ }\bibfield
  {title} {\bibinfo {title} {Charge orders in organic charge-transfer salts},\
  }\href {https://doi.org/10.1088/1367-2630/aa887b} {\bibfield  {journal}
  {\bibinfo  {journal} {New J. Phys.}\ }\textbf {\bibinfo {volume} {19}},\
  \bibinfo {pages} {103033} (\bibinfo {year} {2017})}\BibitemShut {NoStop}%
\bibitem [{\citenamefont {Priya}\ \emph {et~al.}(2026)\citenamefont {Priya},
  \citenamefont {Dressel}, \citenamefont {Liebman},\ and\ \citenamefont
  {Drichko}}]{priya26}%
  \BibitemOpen
  \bibfield  {author} {\bibinfo {author} {\bibfnamefont {S.}~\bibnamefont
  {Priya}}, \bibinfo {author} {\bibfnamefont {M.}~\bibnamefont {Dressel}},
  \bibinfo {author} {\bibfnamefont {J.}~\bibnamefont {Liebman}},\ and\ \bibinfo
  {author} {\bibfnamefont {N.}~\bibnamefont {Drichko}},\ }\bibfield  {title}
  {\bibinfo {title} {Charge-sensitive vibrational modes in {BEDT-TTF} salts:
  Signatures of charge ordering and site charge},\ }\bibfield  {journal}
  {\bibinfo  {journal} {J. Chem. Phys.}\ }\textbf {\bibinfo {volume} {165}},\
  \href {https://doi.org/10.1063/5.0343156} {10.1063/5.0343156} (\bibinfo
  {year} {2026})\BibitemShut {NoStop}%
\bibitem [{\citenamefont {Ilakovac}\ \emph {et~al.}(2017)\citenamefont
  {Ilakovac}, \citenamefont {Carniato}, \citenamefont {Foury-Leylekian},
  \citenamefont {Tomić}, \citenamefont {Pouget}, \citenamefont {Lazić},
  \citenamefont {Joly}, \citenamefont {Miyagawa}, \citenamefont {Kanoda},\ and\
  \citenamefont {Nicolaou}}]{ilakovac17}%
  \BibitemOpen
  \bibfield  {author} {\bibinfo {author} {\bibfnamefont {V.}~\bibnamefont
  {Ilakovac}}, \bibinfo {author} {\bibfnamefont {S.}~\bibnamefont {Carniato}},
  \bibinfo {author} {\bibfnamefont {P.}~\bibnamefont {Foury-Leylekian}},
  \bibinfo {author} {\bibfnamefont {S.}~\bibnamefont {Tomić}}, \bibinfo
  {author} {\bibfnamefont {J.-P.}\ \bibnamefont {Pouget}}, \bibinfo {author}
  {\bibfnamefont {P.}~\bibnamefont {Lazić}}, \bibinfo {author} {\bibfnamefont
  {Y.}~\bibnamefont {Joly}}, \bibinfo {author} {\bibfnamefont {K.}~\bibnamefont
  {Miyagawa}}, \bibinfo {author} {\bibfnamefont {K.}~\bibnamefont {Kanoda}},\
  and\ \bibinfo {author} {\bibfnamefont {A.}~\bibnamefont {Nicolaou}},\
  }\bibfield  {title} {\bibinfo {title} {Resonant inelastic x-ray scattering
  probes the electron-phonon coupling in the spin liquid
  $\kappa$-({BEDT-TTF})\textsubscript{2}{Cu}\textsubscript{2}({CN})\textsubscript{3}},\
  }\href {https://doi.org/10.1103/physrevb.96.184303} {\bibfield  {journal}
  {\bibinfo  {journal} {Phys. Rev. B}\ }\textbf {\bibinfo {volume} {96}},\
  \bibinfo {pages} {184303} (\bibinfo {year} {2017})}\BibitemShut {NoStop}%
\bibitem [{\citenamefont {Pustogow}(2022)}]{pustogow22}%
  \BibitemOpen
  \bibfield  {author} {\bibinfo {author} {\bibfnamefont {A.}~\bibnamefont
  {Pustogow}},\ }\bibfield  {title} {\bibinfo {title} {Thirty-year anniversary
  of
  $\kappa$-({BEDT-TTF})\textsubscript{2}{Cu}\textsubscript{2}({CN})\textsubscript{3}:
  Reconciling the spin gap in a spin-liquid candidate},\ }\href
  {https://doi.org/10.3390/solids3010007} {\bibfield  {journal} {\bibinfo
  {journal} {Solids}\ }\textbf {\bibinfo {volume} {3}},\ \bibinfo {pages} {93}
  (\bibinfo {year} {2022})}\BibitemShut {NoStop}%
\bibitem [{\citenamefont {Liebman}\ \emph {et~al.}(2024)\citenamefont
  {Liebman}, \citenamefont {Miyagawa}, \citenamefont {Kanoda},\ and\
  \citenamefont {Drichko}}]{liebman24}%
  \BibitemOpen
  \bibfield  {author} {\bibinfo {author} {\bibfnamefont {J.}~\bibnamefont
  {Liebman}}, \bibinfo {author} {\bibfnamefont {K.}~\bibnamefont {Miyagawa}},
  \bibinfo {author} {\bibfnamefont {K.}~\bibnamefont {Kanoda}},\ and\ \bibinfo
  {author} {\bibfnamefont {N.}~\bibnamefont {Drichko}},\ }\bibfield  {title}
  {\bibinfo {title} {Novel dipole-lattice coupling in the quantum spin liquid
  material
  {BEDT-TTF}\textsubscript{2}{Cu}\textsubscript{2}({CN})\textsubscript{3}},\
  }\href {https://doi.org/10.1103/physrevb.110.165105} {\bibfield  {journal}
  {\bibinfo  {journal} {Phys. Rev. B}\ }\textbf {\bibinfo {volume} {110}},\
  \bibinfo {pages} {165105} (\bibinfo {year} {2024})}\BibitemShut {NoStop}%
\bibitem [{\citenamefont {Sedlmeier}\ \emph {et~al.}(2012)\citenamefont
  {Sedlmeier}, \citenamefont {Els\"asser}, \citenamefont {Neubauer},
  \citenamefont {Beyer}, \citenamefont {Wu}, \citenamefont {Ivek},
  \citenamefont {Tomi\'c}, \citenamefont {Schlueter},\ and\ \citenamefont
  {Dressel}}]{sedlmeier12}%
  \BibitemOpen
  \bibfield  {author} {\bibinfo {author} {\bibfnamefont {K.}~\bibnamefont
  {Sedlmeier}}, \bibinfo {author} {\bibfnamefont {S.}~\bibnamefont
  {Els\"asser}}, \bibinfo {author} {\bibfnamefont {D.}~\bibnamefont
  {Neubauer}}, \bibinfo {author} {\bibfnamefont {R.}~\bibnamefont {Beyer}},
  \bibinfo {author} {\bibfnamefont {D.}~\bibnamefont {Wu}}, \bibinfo {author}
  {\bibfnamefont {T.}~\bibnamefont {Ivek}}, \bibinfo {author} {\bibfnamefont
  {S.}~\bibnamefont {Tomi\'c}}, \bibinfo {author} {\bibfnamefont {J.~A.}\
  \bibnamefont {Schlueter}},\ and\ \bibinfo {author} {\bibfnamefont
  {M.}~\bibnamefont {Dressel}},\ }\bibfield  {title} {\bibinfo {title} {Absence
  of charge order in the dimerized $\kappa$-phase {BEDT}-{TTF} salts},\ }\href
  {https://doi.org/10.1103/PhysRevB.86.245103} {\bibfield  {journal} {\bibinfo
  {journal} {Phys. Rev. B}\ }\textbf {\bibinfo {volume} {86}},\ \bibinfo
  {pages} {245103} (\bibinfo {year} {2012})}\BibitemShut {NoStop}%
\bibitem [{\citenamefont {Yakushi}\ \emph {et~al.}(2015)\citenamefont
  {Yakushi}, \citenamefont {Yamamoto}, \citenamefont {Yamamoto}, \citenamefont
  {Saito},\ and\ \citenamefont {Kawamoto}}]{yakushi15}%
  \BibitemOpen
  \bibfield  {author} {\bibinfo {author} {\bibfnamefont {K.}~\bibnamefont
  {Yakushi}}, \bibinfo {author} {\bibfnamefont {K.}~\bibnamefont {Yamamoto}},
  \bibinfo {author} {\bibfnamefont {T.}~\bibnamefont {Yamamoto}}, \bibinfo
  {author} {\bibfnamefont {Y.}~\bibnamefont {Saito}},\ and\ \bibinfo {author}
  {\bibfnamefont {A.}~\bibnamefont {Kawamoto}},\ }\bibfield  {title} {\bibinfo
  {title} {Raman spectroscopy study of charge fluctuation in the spin-liquid
  candidate
  $\kappa$-({BEDT-TTF})\textsubscript{2}{Cu}\textsubscript{2}({CN})\textsubscript{3}},\
  }\href {https://doi.org/10.7566/jpsj.84.084711} {\bibfield  {journal}
  {\bibinfo  {journal} {J. Phys. Soc. Jpn.}\ }\textbf {\bibinfo {volume}
  {84}},\ \bibinfo {pages} {084711} (\bibinfo {year} {2015})}\BibitemShut
  {NoStop}%
\bibitem [{\citenamefont {Foury-Leylekian}\ \emph {et~al.}(2018)\citenamefont
  {Foury-Leylekian}, \citenamefont {Ilakovac}, \citenamefont {Bal{\'{e}}dent},
  \citenamefont {Fertey}, \citenamefont {Arakcheeva}, \citenamefont {Milat},
  \citenamefont {Petermann}, \citenamefont {Guillier}, \citenamefont
  {Miyagawa}, \citenamefont {Kanoda}, \citenamefont {Alemany}, \citenamefont
  {Canadell}, \citenamefont {Tomic},\ and\ \citenamefont
  {Pouget}}]{foury-leylekian18}%
  \BibitemOpen
  \bibfield  {author} {\bibinfo {author} {\bibfnamefont {P.}~\bibnamefont
  {Foury-Leylekian}}, \bibinfo {author} {\bibfnamefont {V.}~\bibnamefont
  {Ilakovac}}, \bibinfo {author} {\bibfnamefont {V.}~\bibnamefont
  {Bal{\'{e}}dent}}, \bibinfo {author} {\bibfnamefont {P.}~\bibnamefont
  {Fertey}}, \bibinfo {author} {\bibfnamefont {A.}~\bibnamefont {Arakcheeva}},
  \bibinfo {author} {\bibfnamefont {O.}~\bibnamefont {Milat}}, \bibinfo
  {author} {\bibfnamefont {D.}~\bibnamefont {Petermann}}, \bibinfo {author}
  {\bibfnamefont {G.}~\bibnamefont {Guillier}}, \bibinfo {author}
  {\bibfnamefont {K.}~\bibnamefont {Miyagawa}}, \bibinfo {author}
  {\bibfnamefont {K.}~\bibnamefont {Kanoda}}, \bibinfo {author} {\bibfnamefont
  {P.}~\bibnamefont {Alemany}}, \bibinfo {author} {\bibfnamefont
  {E.}~\bibnamefont {Canadell}}, \bibinfo {author} {\bibfnamefont
  {S.}~\bibnamefont {Tomic}},\ and\ \bibinfo {author} {\bibfnamefont {J.-P.}\
  \bibnamefont {Pouget}},\ }\bibfield  {title} {\bibinfo {title}
  {({BEDT}-{TTF})\textsubscript{2}{Cu}\textsubscript{2}({CN})\textsubscript{3}
  spin liquid: Beyond the average structure},\ }\href
  {https://doi.org/10.3390/cryst8040158} {\bibfield  {journal} {\bibinfo
  {journal} {Crystals}\ }\textbf {\bibinfo {volume} {8}},\ \bibinfo {pages}
  {158} (\bibinfo {year} {2018})}\BibitemShut {NoStop}%
\bibitem [{\citenamefont {Bayo}\ \emph {et~al.}(2025)\citenamefont {Bayo},
  \citenamefont {Balédent}, \citenamefont {Boccuni}, \citenamefont {Ilakovac},
  \citenamefont {Carniato}, \citenamefont {Foury-Leylekian}, \citenamefont
  {Pouget}, \citenamefont {Nicolaou}, \citenamefont {Ruotsalainen},
  \citenamefont {Bendounan}, \citenamefont {Bouaziz}, \citenamefont {Joly},
  \citenamefont {Miyagawa}, \citenamefont {Kanoda},\ and\ \citenamefont
  {Tomić}}]{bayo25}%
  \BibitemOpen
  \bibfield  {author} {\bibinfo {author} {\bibfnamefont {D.}~\bibnamefont
  {Bayo}}, \bibinfo {author} {\bibfnamefont {V.}~\bibnamefont {Balédent}},
  \bibinfo {author} {\bibfnamefont {A.}~\bibnamefont {Boccuni}}, \bibinfo
  {author} {\bibfnamefont {V.}~\bibnamefont {Ilakovac}}, \bibinfo {author}
  {\bibfnamefont {S.}~\bibnamefont {Carniato}}, \bibinfo {author}
  {\bibfnamefont {P.}~\bibnamefont {Foury-Leylekian}}, \bibinfo {author}
  {\bibfnamefont {J.-P.}\ \bibnamefont {Pouget}}, \bibinfo {author}
  {\bibfnamefont {A.}~\bibnamefont {Nicolaou}}, \bibinfo {author}
  {\bibfnamefont {K.}~\bibnamefont {Ruotsalainen}}, \bibinfo {author}
  {\bibfnamefont {A.}~\bibnamefont {Bendounan}}, \bibinfo {author}
  {\bibfnamefont {M.}~\bibnamefont {Bouaziz}}, \bibinfo {author} {\bibfnamefont
  {Y.}~\bibnamefont {Joly}}, \bibinfo {author} {\bibfnamefont {K.}~\bibnamefont
  {Miyagawa}}, \bibinfo {author} {\bibfnamefont {K.}~\bibnamefont {Kanoda}},\
  and\ \bibinfo {author} {\bibfnamefont {S.}~\bibnamefont {Tomić}},\
  }\bibfield  {title} {\bibinfo {title} {Highly variable carbon environment in
  the
  $\kappa$-({BEDT-TTF})\textsubscript{2}{Cu}\textsubscript{2}({CN})\textsubscript{3}
  salt probed by carbon {K}--edge x-ray absorption and resonant inelastic x-ray
  scattering spectroscopy},\ }\href
  {https://doi.org/10.1103/physrevb.111.125160} {\bibfield  {journal} {\bibinfo
   {journal} {Phys. Rev. B}\ }\textbf {\bibinfo {volume} {111}},\ \bibinfo
  {pages} {125160} (\bibinfo {year} {2025})}\BibitemShut {NoStop}%
\bibitem [{\citenamefont {Galimzyanov}\ \emph {et~al.}(1989)\citenamefont
  {Galimzyanov}, \citenamefont {Ignat’ev}, \citenamefont {Kushch},
  \citenamefont {Laukhin}, \citenamefont {Makova}, \citenamefont {Merzhanov},
  \citenamefont {Rozenberg}, \citenamefont {Shibaeva},\ and\ \citenamefont
  {Yagubskii}}]{galimzyanov89}%
  \BibitemOpen
  \bibfield  {author} {\bibinfo {author} {\bibfnamefont {A.}~\bibnamefont
  {Galimzyanov}}, \bibinfo {author} {\bibfnamefont {A.}~\bibnamefont
  {Ignat’ev}}, \bibinfo {author} {\bibfnamefont {N.}~\bibnamefont {Kushch}},
  \bibinfo {author} {\bibfnamefont {V.}~\bibnamefont {Laukhin}}, \bibinfo
  {author} {\bibfnamefont {M.}~\bibnamefont {Makova}}, \bibinfo {author}
  {\bibfnamefont {V.}~\bibnamefont {Merzhanov}}, \bibinfo {author}
  {\bibfnamefont {L.}~\bibnamefont {Rozenberg}}, \bibinfo {author}
  {\bibfnamefont {R.}~\bibnamefont {Shibaeva}},\ and\ \bibinfo {author}
  {\bibfnamefont {E.}~\bibnamefont {Yagubskii}},\ }\bibfield  {title} {\bibinfo
  {title} {New cation-radical salts of {BEDT-TTF} with the octahedral metal
  complex anions {PtCl}\textsubscript{6}\textsuperscript{2-},
  {PtBr}\textsubscript{6}\textsuperscript{2-},
  {TeCl}\textsubscript{6}\textsuperscript{2-} and
  {SnCl}\textsubscript{6}\textsuperscript{2-}. {S}ynthesis, structure and
  properties},\ }\href {https://doi.org/10.1016/0379-6779(89)90859-X}
  {\bibfield  {journal} {\bibinfo  {journal} {Synthetic Met.}\ }\textbf
  {\bibinfo {volume} {33}},\ \bibinfo {pages} {81} (\bibinfo {year}
  {1989})}\BibitemShut {NoStop}%
\bibitem [{\citenamefont {Świetlik}\ \emph {et~al.}(2001)\citenamefont
  {Świetlik}, \citenamefont {Połomska}, \citenamefont {Ouahab},\ and\
  \citenamefont {Guillevic}}]{swietlik01}%
  \BibitemOpen
  \bibfield  {author} {\bibinfo {author} {\bibfnamefont {R.}~\bibnamefont
  {Świetlik}}, \bibinfo {author} {\bibfnamefont {M.}~\bibnamefont
  {Połomska}}, \bibinfo {author} {\bibfnamefont {L.}~\bibnamefont {Ouahab}},\
  and\ \bibinfo {author} {\bibfnamefont {J.}~\bibnamefont {Guillevic}},\
  }\bibfield  {title} {\bibinfo {title} {Infrared and {R}aman studies of the
  $\kappa$-phase charge-transfer salts formed by {BEDT-TTF} and magnetic anions
  {M(CN)}\textsubscript{6}\textsuperscript{3-} (where {M} = {CoIII}, {FeIII},
  {CrIII})},\ }\href {https://doi.org/10.1039/b007752f} {\bibfield  {journal}
  {\bibinfo  {journal} {J. Mater. Chem.}\ }\textbf {\bibinfo {volume} {11}},\
  \bibinfo {pages} {1313} (\bibinfo {year} {2001})}\BibitemShut {NoStop}%
\bibitem [{\citenamefont {Świetlik}\ \emph {et~al.}(2004)\citenamefont
  {Świetlik}, \citenamefont {Ouahab}, \citenamefont {Guillevic},\ and\
  \citenamefont {Yakushi}}]{swietlik04}%
  \BibitemOpen
  \bibfield  {author} {\bibinfo {author} {\bibfnamefont {R.}~\bibnamefont
  {Świetlik}}, \bibinfo {author} {\bibfnamefont {L.}~\bibnamefont {Ouahab}},
  \bibinfo {author} {\bibfnamefont {J.}~\bibnamefont {Guillevic}},\ and\
  \bibinfo {author} {\bibfnamefont {K.}~\bibnamefont {Yakushi}},\ }\bibfield
  {title} {\bibinfo {title} {Infrared and {R}aman studies of the charge
  ordering in the organic semiconductor
  $\kappa$‐[{E}t\textsubscript{4}{N}][{Co(CN)}\textsubscript{6}({ET})\textsubscript{4}]·3{H}\textsubscript{2}{O}},\
  }\href {https://doi.org/10.1002/masy.200450821} {\bibfield  {journal}
  {\bibinfo  {journal} {Macromol. Symp.}\ }\textbf {\bibinfo {volume} {212}},\
  \bibinfo {pages} {219} (\bibinfo {year} {2004})}\BibitemShut {NoStop}%
\bibitem [{\citenamefont {{\'{S}}wietlik}\ \emph {et~al.}(2006)\citenamefont
  {{\'{S}}wietlik}, \citenamefont {{\L}api{\'{n}}ski}, \citenamefont
  {Po{\l}omska}, \citenamefont {Ouahab},\ and\ \citenamefont
  {Ota}}]{swietlik06}%
  \BibitemOpen
  \bibfield  {author} {\bibinfo {author} {\bibfnamefont {R.}~\bibnamefont
  {{\'{S}}wietlik}}, \bibinfo {author} {\bibfnamefont {A.}~\bibnamefont
  {{\L}api{\'{n}}ski}}, \bibinfo {author} {\bibfnamefont {M.}~\bibnamefont
  {Po{\l}omska}}, \bibinfo {author} {\bibfnamefont {L.}~\bibnamefont
  {Ouahab}},\ and\ \bibinfo {author} {\bibfnamefont {A.}~\bibnamefont {Ota}},\
  }\bibfield  {title} {\bibinfo {title} {Infrared and {R}aman investigations of
  the charge ordering phenomena in the monoclinic salts
  $\kappa$-({BEDT}-{TTF})\textsubscript{4}[{M}({CN})\textsubscript{6}][{N}({C}\textsubscript{2}{H}\textsubscript{5})\textsubscript{4}]\textsubscript{2}{H}\textsubscript{2}{O}},\
  }\href {https://doi.org/10.1007/s10909-006-9108-9} {\bibfield  {journal}
  {\bibinfo  {journal} {J. Low Temp. Phys.}\ }\textbf {\bibinfo {volume}
  {142}},\ \bibinfo {pages} {653} (\bibinfo {year} {2006})}\BibitemShut
  {NoStop}%
\bibitem [{\citenamefont {Ota}\ \emph {et~al.}(2007)\citenamefont {Ota},
  \citenamefont {Ouahab}, \citenamefont {Golhen}, \citenamefont {Yoshida},
  \citenamefont {Maesato}, \citenamefont {Saito},\ and\ \citenamefont
  {{\'{S}}wietlik}}]{ota07}%
  \BibitemOpen
  \bibfield  {author} {\bibinfo {author} {\bibfnamefont {A.}~\bibnamefont
  {Ota}}, \bibinfo {author} {\bibfnamefont {L.}~\bibnamefont {Ouahab}},
  \bibinfo {author} {\bibfnamefont {S.}~\bibnamefont {Golhen}}, \bibinfo
  {author} {\bibfnamefont {Y.}~\bibnamefont {Yoshida}}, \bibinfo {author}
  {\bibfnamefont {M.}~\bibnamefont {Maesato}}, \bibinfo {author} {\bibfnamefont
  {G.}~\bibnamefont {Saito}},\ and\ \bibinfo {author} {\bibfnamefont
  {R.}~\bibnamefont {{\'{S}}wietlik}},\ }\bibfield  {title} {\bibinfo {title}
  {Phase transition from {M}ott insulating phase into the charge ordering phase
  with molecular deformation in charge-transfer salts
  $\kappa$-({ET})\textsubscript{4}[{M(CN)}\textsubscript{6}][{N(C}\textsubscript{2}{H}\textsubscript{5})\textsubscript{4}]·2{H}\textsubscript{2}{O}
  ({M} = {Co}\textsuperscript{III} and {Fe}\textsuperscript{III})},\ }\href
  {https://doi.org/10.1021/cm063036k} {\bibfield  {journal} {\bibinfo
  {journal} {Chem. Mater.}\ }\textbf {\bibinfo {volume} {19}},\ \bibinfo
  {pages} {2455} (\bibinfo {year} {2007})}\BibitemShut {NoStop}%
\bibitem [{\citenamefont {Bardin}\ \emph {et~al.}(2023)\citenamefont {Bardin},
  \citenamefont {Prokhorova},\ and\ \citenamefont {Buravov}}]{bardin23}%
  \BibitemOpen
  \bibfield  {author} {\bibinfo {author} {\bibfnamefont {A.~A.}\ \bibnamefont
  {Bardin}}, \bibinfo {author} {\bibfnamefont {T.~G.}\ \bibnamefont
  {Prokhorova}},\ and\ \bibinfo {author} {\bibfnamefont {L.~I.}\ \bibnamefont
  {Buravov}},\ }\bibfield  {title} {\bibinfo {title} {A new charge-ordered
  molecular conductor:
  $\kappa$-({BEDT-TTF})\textsubscript{2}{K}\textsuperscript{+}(18-crown-6)[{Co}\textsuperscript{II}{(NCS)}\textsubscript{4}]·({H}\textsubscript{2}{O})},\
  }\href {https://doi.org/10.3390/cryst13101504} {\bibfield  {journal}
  {\bibinfo  {journal} {Crystals}\ }\textbf {\bibinfo {volume} {13}},\ \bibinfo
  {pages} {1504} (\bibinfo {year} {2023})}\BibitemShut {NoStop}%
\bibitem [{\citenamefont {Watanabe}\ \emph {et~al.}(2004)\citenamefont
  {Watanabe}, \citenamefont {Noda}, \citenamefont {Nogami},\ and\ \citenamefont
  {Mori}}]{watanabe04}%
  \BibitemOpen
  \bibfield  {author} {\bibinfo {author} {\bibfnamefont {M.}~\bibnamefont
  {Watanabe}}, \bibinfo {author} {\bibfnamefont {Y.}~\bibnamefont {Noda}},
  \bibinfo {author} {\bibfnamefont {Y.}~\bibnamefont {Nogami}},\ and\ \bibinfo
  {author} {\bibfnamefont {H.}~\bibnamefont {Mori}},\ }\bibfield  {title}
  {\bibinfo {title} {Transfer integrals and the spatial pattern of charge
  ordering in
  $\theta$-({BEDT-TTF})\textsubscript{2}{RbZn}({SCN})\textsubscript{4} at 90
  {K}},\ }\href {https://doi.org/10.1143/JPSJ.73.116} {\bibfield  {journal}
  {\bibinfo  {journal} {J. Phys. Soc. Jpn.}\ }\textbf {\bibinfo {volume}
  {73}},\ \bibinfo {pages} {116} (\bibinfo {year} {2004})}\BibitemShut
  {NoStop}%
\bibitem [{\citenamefont {Takahashi}\ \emph {et~al.}(2006)\citenamefont
  {Takahashi}, \citenamefont {Nogami},\ and\ \citenamefont
  {Yakushi}}]{takahashi06}%
  \BibitemOpen
  \bibfield  {author} {\bibinfo {author} {\bibfnamefont {T.}~\bibnamefont
  {Takahashi}}, \bibinfo {author} {\bibfnamefont {Y.}~\bibnamefont {Nogami}},\
  and\ \bibinfo {author} {\bibfnamefont {K.}~\bibnamefont {Yakushi}},\
  }\bibfield  {title} {\bibinfo {title} {Charge ordering in organic
  conductors},\ }\href {https://doi.org/10.1143/JPSJ.75.051008} {\bibfield
  {journal} {\bibinfo  {journal} {J. Phys. Soc. Jpn.}\ }\textbf {\bibinfo
  {volume} {75}},\ \bibinfo {pages} {051008} (\bibinfo {year}
  {2006})}\BibitemShut {NoStop}%
\end{thebibliography}
\let\clearpage\relax

\end{document}